\documentclass[a4paper,USenglish,cleveref,autoref]{lipics-v2021}

\usepackage[utf8]{inputenc}
\usepackage{amssymb}
\usepackage{bm} %
\usepackage{graphicx}
\usepackage{caption}
\usepackage{enumerate}
\usepackage{relsize}
\usepackage{color}
\usepackage{verbatim}
\usepackage{amsthm}
\usepackage[ruled,linesnumbered]{algorithm2e}
\usepackage{xr}
\usepackage{tikz-cd}
\usepackage{etoolbox}
\usepackage{float}
\usepackage{xspace}
\usepackage{faktor}
\usepackage{yhmath}
\usepackage{wrapfig}
\restylefloat{figure}

\usepackage{epsfig}
\usepackage{url} 
\usepackage{amsmath}
\usepackage{tikz}
\usepackage[all,knot,poly]{xy}
\usetikzlibrary{matrix}
\usepackage{hyperref}
\usepackage{pst-optic}
\usepackage{graphicx}
\usetikzlibrary{calc}

\nolinenumbers

\input{macrosDUAL}

\newcommand{\interior}[2][]{{(#2)^{o#1}}}

\title{Stone Duality Proofs for Colorless Distributed Computability Theorems}

\author{Cameron Calk}{Aix-Marseille University \& CNRS UMR7020, Marseille, France}{}{}{}
\author{Emmanuel Godard}{Aix Marseille University \& CNRS UMR7020, Marseille, France}{}{}{}

\authorrunning{C. Calk and E. Godard}

\Copyright{Calk-Godard}
  


\EventEditors{}
\EventNoEds{1}
\EventLongTitle{}
\EventShortTitle{}
\EventAcronym{TBA}
\EventYear{2025}
\hideLIPIcs

\begin{document}
\maketitle

\begin{abstract}

Twenty years ago, Herlihy/Shavit and Saks/Zaharoglou won the Gödel
prize for the introduction of a simplicial semantics for distributed
computing. This line of work culminated in a
characterization of the distributed tasks which can be solved by asynchronous wait-free systems,
resulting in the Asynchronous Computability Theorem (ACT).
In this paper, we extend this semantics
by identifying spectral topology as the natural generalization of the
finite combinatorial topology they employed. In particular, we extend
the topological approach of ACT to any round-based, content-neutral,
full-information protocol. This family of protocols contains the Iterated Immediate Snapshot model ($IIS$),
to which many distributed computation models can be reduced.
In this sense, our work provides first
steps towards a unified topological framework for distributed
computing.

The main insight of this work is in considering global states obtained
after finite executions of a distributed protocol not as abstract
simplicial complexes as was previously done, but as finite spectral spaces, considering the Alexandrov topology on the
associated face posets. Using this point-set topological approach,
coupled with the interpretation of a distributed protocol as an endofunctor $\Pi$
on the category of simplicial complexes, we show that any initial
configuration $\mathcal{I}$ can be associated to a projective limit
system of finite complexes. The limit thereof is a spectral space
$\Pi^\infty(\mathcal{I})$ which precisely encodes the
behavior of the protocol presented by $\Pi$. This leads us to derive
a new general distributed computability theorem using Stone duality: a
protocol $\Pi$ solves a colorless task $(\mathcal{I},\mathcal{O},\Delta)$ if and
only if there exists a spectral map
$f:\Pi^\infty(\mathcal{I})\rightarrow\mathcal{O}$ compatible with
$\Delta$.
  
From this general characterization, we derive known colorless
computability theorems,
and provide new insights into the previously established
connection between task-solvability and continuous maps between
geometric realizations. This is achieved through Stone duality, a well established tool for such tight correspondences in computer science.

\end{abstract}
\paragraph*{Acknowledgements.} 

This work is supported by grant DisQC ANR-22-CE47-0002-01 from the
French National Research Agency, the Amidex fondation and project
ANR-23-PECL-0009 TRUSTINCloudS (PEPR Cloud).
\newpage
\section{Introduction}

One of the most important results in distributed computing is the identification of finite combinatorial topology, in the form of simplicial complexes and their structure maps, as the natural setting of the subject.
A typical distributed problem, a \emph{task}, is for example presented as a triple $(\Ic,\Oc,\Delta)$ where $\Ic$ and $\Oc$ are simplicial complexes, and $\Delta$ a relation between them (see Section~\ref{S:Colorlesstask} for details).
Furthermore, at least for some tasks, it is known that continuous maps between the geometric realizations of associated simplicial complexes classify which tasks can be solved by the protocol. While many generalizations and applications of this topological approach have been developed, still to this day, the direct connection between distributed computing and continuous maps between geometric realizations was unclear. Here, we identify \emph{spectral spaces}, arising as limits of the finite combinatorial topology models of distributed computing, as the underlying phenomenon behind this remarkable connection.
Our work shows that this topological characterization is in fact a result of Stone duality, which is a well established tool for such correspondences in computer science. 
This additionally allows us to obtain a version of the original result which applies to a much wider class of protocols.

\paragraph*{Topological methods for distributed computability}

In a distributed system, several processes attempt to coordinate through some form of communication in order to solve a task. 
As opposed to sequential systems, in which computability can be characterized by various equivalent abstract objects such as Turing machines or $\lambda$-calculus, task-solvability, \ie (distributed) computability, has no known universal model. Moreover, distributed computability is not, in essence, limited by the computing power of each process, the main obstruction being the lack of global information available to each process. Characterizing task-solvability for general models of communication has been a central question in distributed computing since its foundation. In $2004$, Herlihy and Shavit~\cite{HS99}, as well as Saks and Zaharoglou \cite{SZ}, building on the work of Borowsky and Gafni \cite{BGatomicsnapshot,BGequivIIS}, were awarded the Gödel prize for showing that central problems in distributed computability can be understood and solved using topology. 

Indeed, the possible epistemic states of a distributed system fit nicely into finite combinatorial objects called simplicial complexes, which have a clear topological interpretation. 
Concretely, such an object consists of all possible global states of the system{, glued along shared subsets}. Each global state in a distributed system is a collection of local states. This object not only represents the combinatorics of possible configurations, but also encodes the lack of global information available to each process: a process in a certain local state cannot distinguish between two global states if its local state lies in their intersection. Simplicial complexes, which can also be thought of as spaces built by gluing together points, lines, triangles and their higher dimensional analogues, thus precisely encode the epistemic ambiguities of the system.

In sequential computing, a program specification consists of a relation between a set of inputs and a set of outputs. A distributed specification is encoded similarly, but due to the multiplicity of processes, must relate a complex of inputs to a complex of outputs. In the simplicial semantics of distributed computing mentioned above, a \emph{colorless} distributed task is presented as a triple $(\In,\Out,\Delta)$ where \In and \Out are simplicial complexes, and $\Delta$ is a relation between global states in \In and legal output states in \Out which respects the simplicial structure. Classically, global states in distributed systems are given by sets of pairs $(p,v)$ where $p$ is a process and $v$ is a value. However, for certain tasks, this information is superfluous. Indeed, a distributed task is said to be colorless if its specification does not depend of the multiplicity
of values in the inputs and outputs. This is a sub-class of interest that contains agreement tasks, like Consensus.
Such agreement tasks are central to distributed computing since they correspond to reliable recovery from faults in replicated databases, an area of tremendous practical interest. Formally, this means that we can describe such tasks with (achromatic) simplicial complexes, rather than with \emph{chromatic} complexes, in which process identities are encoded by a labeling function, see Section~\ref{SS:CombSemantics}. 

On the other hand, a distributed protocol describes the modalities of exchange of information, for example whether it be via message-passing or a shared-memory object, if it is round-based or asynchronous, or whether processes can crash. Exchanges of information lead to new global states reachable from the input states. These reachable global states, glued along common subsets of processes' local states, form the so-called \emph{protocol complex} $\mathcal P$. As in the case of distributed tasks, distributed protocols can thus also be encoded as a relation $\pi$ between global states in $\In$ and the corresponding reachable states in $\mathcal P$. {In this paper, we consider \emph{achromatic} models of protocols. In cases where the communication does not depend on processes' identities, but only on their values, we can, as in the case of colorless tasks, describe the epistemic dynamic of the communication purely in terms of (achromatic) simplicial complexes.}

A distributed protocol \emph{solves} a task if, for every execution, each process can locally decide on an output value consistent with specification $\Delta$.
In this case we say that the task is \emph{solvable} or, equivalently, \emph{computable}.
For colorless tasks and achromatic protocols, solving corresponds to the existence of a simplicial map $\delta : \Pc \fl \Out$, called the \emph{decision map}, which is compatible with the specification $\Delta$. The simplicial semantics of distributed computing is further detailed in Sections~\ref{S:OFIFunc} and~\ref{SS:DistTaskCat}. 
This combinatorial presentation of distributed systems resulted in many novel impossibility results for colorless task-solvability using purely topological arguments. In particular, for the Iterated Immediate Snapshot ($IIS$) model~\cite{HS99,HKRbook}, colorless task-solvability is related to the existence of a continuous map between the
geometric realizations of the associated input and output complexes. In this paper, we present novel topological and categorical techniques which extend this characterization of {colorless} task-solvability to a wider class of models, while also explaining the phenomena behind the original result.

\paragraph*{Our Contributions}

In short, our main contribution is the identification of spectral topology as a natural and fruitful extension of the simplicial semantics for distributed computing. This perspective allows us to generalize known colorless topological computability results to any round-based, content-neutral, full-information model of computation. We achieve this by encoding such a protocol as an endofunctor $\Pi$ on the category of simplicial complexes. Using this, we associate a spectral space $\speclim\In$ to any input complex $\In$, which can be described abstractly in terms of a projective limit, but also concretely as a space of sequences, the latter being closely related to executions and corresponding knowledge, see Proposition~\ref{P:ProtSpecSpace}. These spaces characterize computability, as is shown in Theorem~\ref{thm-carac}, which states that a protocol $\Pi$ solves a task $(\In, \Out, \Delta)$ if, and only if, there exists a spectral map from $\speclim\In$ to the output complex $\Out$ which respects the specification $\Delta$.

Finally, we show in Theorems~\ref{Theorem1}, \ref{Theorem2} and \ref{Theorem3} that this spectral approach is consistent with the original results for the $IIS$ protocol: a colorless task $(\In, \Out, \Delta)$ is solvable by the $IIS$ protocol if, and only if, there exists a continuous map $f: \real \In \fl \real \Out$ compatible with $\Delta$ between the geometric realizations of the input and output complexes. These results exploit the fact that such protocols correspond simplicially to some subdivision of the input complex. Here, we show that this result holds for any protocol defined by a mesh-shrinking subdivision functor. Indeed, for such protocols we show in Lemma~\ref{L:equivShrink} that the existence of a spectral map $\speclim\In \fl \Out$ compatible with $\Delta$ is equivalent to the existence of a continuous map $\real\In \fl \real\Out$ also compatible with $\Delta$.

\paragraph*{Related Works}
Our approach relies on the interpretation of distributed protocols as endofunctors on the category of simplicial complexes.
While concurrency theory has a long history of using categorical methods (\eg \cite{catWinskel,Winskelmodels}), this is unusual in distributed computing.
Following previous works \cite{GMT15}, preprints~\cite{catdist-x,sheafdist-x,sheafdist-ba} have recently deepened a categorical approach to distributed protocols, using algebraic and sheaf theoretic methods respectively. 
While our functorial encoding of protocols, although {limited for the time being to the case of achromatic protocols}, is similar to that described in~\cite{catdist-x}, our use of Stone duality reunites the algebraic and logical point of view they develop with the topological approach using simplicial complexes, see Section~\ref{sec:ccl} for a further discussion. 

Stone duality is an effective tool in theoretical computer science which expresses the link between computational behaviors and specifications. As emphasized in the survey~\cite{dualityCS}, topology is the ``magic ingredient'' which makes this correspondence work. This article demonstrates that Stone duality can also be naturally applied to questions of task-solvability for distributed systems, while extending and explaining the current literature. We believe that leveraging these duality techniques has the potential to transform research in distributed computing by reconciling the topological, combinatorial, algebraic and logical approaches to the subject, with the long-term goal of establishing a unified semantics for distributed computing.
The present work is a step in this direction, in which we focus on the most simple presentation of distributed computability, namely the case of colorless computability for achromatic protocols.

{The reason for this restriction is two-fold. On the one hand, the construction of the spectral space $\speclim\Ic$ relies on Stone duality and the characterization of spectral spaces as projective limits of finite posets. A further investigation into the properties of projective limits of chromatic simplicial complexes %
  is needed in order to extend our techniques to the chromatic case. On the other hand, the semantic relationship between achromatic and chromatic protocols and their task-solvability for colorless or colored tasks is not well understood. In order to achieve our end-goal of a unified semantics, these relationships need to be faithfully modeled within our framework. For this reason, we provide here a full description of our techniques in the simplest paradigm of distributed computability, namely colorless task-solvability for achromatic models. This will allow us to extend these techniques to chromatic protocols and uncolored tasks, and then to colored tasks, from a solid formal basis. A preliminary investigation into colorless task-solvability for chromatic protocols is presented in Appendix~\ref{S:IIS}, see also the discussion in Section~\ref{sec:ccl}.}

\section{Preliminaries}

Let $X$ and $Y$ be sets. Given a map $f : X \fl Y$ we denote by $\bck f$ the inverse image map associated to $f$, \ie $\bck f : \Pow{Y} \fl \Pow X$, where $\Pow X$ is the powerset of $X$. The forward image map associated to $f$ is denoted by $\fwd f : \Pow X \fl \Pow Y$. 
Given an element $x \in X$ and subsets $A \subseteq X$ and $B \subseteq Y$, we write $f(x) \in Y$ to denote the image of $x$, $f[A]$ or $\fwd f [A]$ for the forward image of $A$, and $\bck f [B]$ for the inverse image of $B$. This helps to notationally distinguish at which level we are considering $f$: on the set or on the powerset.

We use standard notation and terminology for order and duality theory, see e.g
\cite{spectralBook,dualityBook} for relevant definitions.  Given a
poset $P$, we denote by $\ulatt P$ and $\dlatt P$ the lattices of
up-sets and down-sets of $P$, respectively. Given a monotone map $f : P\fl Q$ between posets, we will abuse notation by writing $\bck f : \dlatt Q \fl \dlatt P$ and $\fwd f : \dlatt P \fl \dlatt Q$ for the restriction of $\bck f$ and $\fwd f$ to downsets.

\subsection{Simplicial complexes}\label{SS:SimpComp}

For the purposes of this paper, we will consider a \emph{simplicial complex} $\Cr$ over a finite set of vertices $V$ to be a down-set in the poset $(\Pow{V}\setminus\{\emptyset\}, \subseteq)$.
This is a simple reformulation of the classical definition of simplicial complexes. Elements $\sigma \in \Cr$ are called \emph{simplices}, minimal elements are called \emph{vertices} and downsets $A \subseteq \Cr$ are called \emph{subcomplexes}. We denote by $V_\Cr$ the set of vertices of $\Cr$, and by $\dlatt \Cr$ its lattice of subcomplexes. 
We denote by $\dim(\sigma) := |\sigma|-1$ the \emph{dimension} of $\sigma$, and by $\dim(\Cr) = \max\{\dim(\sigma) \mid \sigma\in \Cr\}$ the dimension of a complex $\Cr$. A simplex of dimension $k$ will sometimes be referred to as a $k$-simplex.

A \emph{simplicial map} $f : \Cr \fl \Cr'$ is given by a map $g : V_\Cr \fl V_{\Cr'}$ such that $f = \fwd g$. Note that the dimension of $f(\sigma)$ may be of smaller than that of $\sigma$. We say that $f$ is \emph{rigid} when it preserves dimension. The category of simplicial complexes with simplicial maps is denoted by $\scompcat$. 

In the topological approach to distributed computing, another type of map is often also considered. Given two simplicial complexes $\Cr, \Cr'$, a \emph{carrier map} from $\Cr$ to $\Cr'$ is a map $\phi : \Cr \fl \dlatt{\Cr'}$, such that for $\sigma \subseteq \tau$ in $\Cr$, we have $\phi(\sigma) \subseteq \phi(\tau)$. A carrier map is \emph{rigid} when it preserves dimension, \ie for a simplex $\sigma$ of dimension $d$, $\phi(\sigma)$ is a subcomplex of $\Cr'$ of dimension $d$. We say that it is \emph{strict} if it preserves intersections, \ie $\phi(\sigma \cap \tau) = \phi(\sigma) \cap \phi(\tau)$.
However, note that when $\phi$ is not strict, we still have $\phi(\sigma \cap \tau) \subseteq \phi(\sigma) \cap \phi(\tau)$.

Finally, we endow these structures with a topology. It is well known \cite[Section~2.2]{dualityBook} that finite posets can be endowed with a topology given by up- and down-sets. This is known as the Alexandrov topology. Since simplicial complexes are a special kind of poset, we will consider them as finite topological spaces endowed with this topology.
Explicitly, we consider the topology on $\Cr$ in which open sets are up-sets $U \in \ulatt \Cr$ and closed sets are down-sets $D \in \dlatt\Cr$. With this topology, $\Cr$ is a finite $T_0$ space.

\subsection{Spectral spaces}

Finite $T_0$ spaces are examples of special topological spaces called \emph{spectral} or \emph{Stone spaces}, see \cite{spectralBook} for a very complete presentation of these spaces and associated results. Recall that in a topological space $X$, a subset $A \subseteq X$ is said to be \emph{compact} if any open cover of $A$ can be refined to a finite open cover\footnote{Note that we do not require $X$ to be Hausdorff.}. In non-Hausdorff spaces, a compact set is not necessarily closed. For this reason, in such spaces we may consider the set $\copen X$ of \emph{compact-open} subsets of $X$, \ie those which are compact \emph{and} open.
A $T_0$, compact and sober topological space $X$ is a \emph{spectral space} when $\copen X$ is closed under finite intersections and is in addition a basis for its topology, \ie its open sets are generated under arbitrary unions of elements of $\copen X$. A \emph{spectral} map is a continuous function $f: X \fl Y$ between spectral spaces, such that for every $U \in \copen Y$, $\bck f[U] \in \copen X$. The category of spectral spaces and spectral maps is denoted by $\Spec$.
In the case of a finite spectral space, which corresponds precisely to a poset $P$ endowed with the Alexandrov topology, we have $\copen P = \ulatt P$, and spectral maps are simply order-preserving maps \cite[Section~2.2]{dualityBook}. In other words, the category $\Posf$ of finite posets and order-preserving maps is a full sub-category of $\Spec$.

\subsection{Stone duality}

In this paper we will make extensive use of Stone duality, which establishes a useful connection between distributive lattices and spectral spaces. Given a lattice $L$, recall that a \emph{filter} $F \subseteq L$ is a non-empty up-set which is closed under meets. A filter $P$ is \emph{proper} when $P \neq L$ and is \emph{prime} if it is proper and if, for all $a,b \in L$, $a\vee b \in P$ implies that $a \in P$ or $b\in P$. We denote by $\pfilt L$ the set of prime filters of $L$. Endowing $\pfilt L$ with the topology generated by the sets 
$
\hat a = \{ F \in X \mid a \in F \},
$
for $a\in L$, we obtain a spectral space $\St L$. Denoting by $\DL$ the category of distributive lattices and lattice homomorphisms, and by $\Spec$ the category of spectral spaces and spectral maps, we have a pair of contravariant functors between $\DL$ and $\Spec$:
\[
\St{-} : \op\DL \fl \Spec
\qquad \text{and}\qquad
\copen{-} : \op\Spec \fl \DL
\]
establishing a \emph{duality} between $\DL$ and $\Spec$. Concretely, this means that, given a spectral map $f: X \fl X'$ between spectral spaces, we obtain a lattice homomorphism $\bck f : \copen {X'} \fl \copen X$ given by inverse image. Conversely, a lattice homomorphism $h : L \fl L'$ yields a spectral map $\St h : \St{L'} \fl \St{L}$, and these assignments are mutually inverse. For a more in-depth treatment of Stone duality, see \cite{dualityBook}.

As mentioned above, a finite spectral space $X$ is isomorphic to a poset $P$ endowed with the Alexandrov topology, and we have $\copen X \cong \ulatt P$. Conversely, given a finite distributive lattice $L$, there is a bijection between $\pfilt L$ and the set $\JI L$ of join-prime elements of $L$. Summing this up, in the finite case, Stone duality specializes to a duality between $\Posf$ and $\DLf$, the category of finite distributive lattices. This is called \emph{Birkhoff duality}, see \cite{dualityBook} for more information. Note however that since $\dlatt P$ is order-dual to $\ulatt P$, this duality also applies to down-set lattices.

\section{Functorial approach for full-information distributed protocols}\label{S:OFIFunc}

In this section, we describe the mathematical objects which we use to model distributed protocols. Due to the choice of models we are considering, namely {achromatic, content-neutral,} round-based full-information adversaries, we can reformulate the operational definition of distributed protocols as endofunctors on the category of simplicial complexes.
Before describing this functorial presentation of protocols, we first recall the definitions and intuitions of the simplicial semantics of distributed systems, a full account of which can be found in~\cite{HKRbook}, while a more concise presentation is given in~\cite{DCcolumn}.

\subsection{Simplicial semantics of distributed computing}\label{SS:CombSemantics}
Given some number of processes, a \emph{chromatic distributed system} is the given of a set of possible \emph{local views} or \emph{states} for each process, along with subsets thereof describing \emph{global states}. In the pioneering work of Herlihy and Shavit~\cite{HS99}, the authors established a simplicial semantics for task-solvability. Indeed, considering not only global states, but also their subsets, we obtain a simplicial complex. For example, given two processes $p$ and $q$, each of which can have the possible local state $0$ or $1$,
we obtain the chromatic simplicial complex pictured below on the right, in which the local views of $p$ (resp.~$q$) are colored red (resp.~blue). This encoding of a distributed system thus not only contains all global states, but also encodes their indistinguishability for each local state. Indeed, when {the red process has} value $1$, it cannot tell whether it is in a global state in which blue has $0$ or $1$.
Uncertainty here is encoded simplicially by it being a member of two distinct $1$-simplices.

\begin{wrapfigure}{r}{.2\textwidth}
\begin{center}
\begin{tikzpicture}[scale=1.8, every node/.style={font=\small}]

\node[circle, fill=blue!30, draw] (p0) at (0,0) {$0$};
\node[circle, fill=blue!30, draw] (p1) at (1,1) {$1$};

\node[circle, fill=red!30, draw] (q1) at (1,0) {$1$};
\node[circle, fill=red!30, draw] (q0) at (0,1) {$0$};

\draw (p0) -- (q0);
\draw (p0) -- (q1);
\draw (p1) -- (q0);
\draw (p1) -- (q1);

\node[circle, fill=gray!30, draw] (0) at (0,-0.75) {$0$};
\node[circle, fill=gray!30, draw] (1) at (1,-0.75) {$1$};

\draw (0) -- (1);

\end{tikzpicture}
\end{center}
\end{wrapfigure}

Chromatic distributed systems are thus encoded as simplicial complexes decorated with colors representing process identities. However, in many cases of interest in distributed computing, the specification of which process has which view is unimportant; only the \emph{values} present in the system are considered. The identifiers can thus be projected out of the system, resulting in a regular (achromatic) simplicial complexes. Therefore, in the achromatic case, the interpretation of a simplex $\{v_1, \dots, v_k\}$ is that there exists a global state in which each value $v_i$ is the local view of at least one process in the system. The simplicial complex corresponding to the chromatic example described above is pictured just below it.

\subsection{Colorless Distributed tasks}
\label{S:Colorlesstask}
\begin{wrapfigure}{lx<}{.3\textwidth}
\begin{tikzpicture}[scale=.4]
      \node[] at (-1,2.5) {$\In$};
      \node[circle, fill=gray!30, draw] (0) at (0,5.0) {$0$};
      \node[circle, fill=gray!30, draw] (1) at (0,0) {$1$};
      \draw (0) -- (1);

      \node[] at (8.5,2.5) {$\Out$};
      \node[circle, fill=gray!30, draw] (5) at (7,5.0) {$0$};
      \node[circle, fill=gray!30, draw] (4) at (7,0) {$1$};

      \node[blue] at (3,4) {$\Delta$};
      
      \node[] (12) at (0,2.5) {};
      \node[] (13) at (5.5,2.5) {};      
      \draw[blue,label=north:${\Delta}$] (12) -- (13);
      
      \node[] (14) at (5,2.5) {};
      \node[] (15) at (5,2) {};
      \node[] (16) at (5,3) {};
      \draw[->,blue] (16) to[bend right] (4);
      \draw[->,blue] (15) to[bend left] (5);
      \draw[->,blue] (0) -- (5);
      \draw[->,blue] (1) -- (4);

\end{tikzpicture}
\end{wrapfigure}
In order to provide a faithful semantics of task-solvability, we also simplicially describe tasks, \ie distributed program specifications. For a sequential program, a specification is simply a relation between a set of input values and a set of output values.
Since a distributed algorithm outputs a value at each process, a distributed task must relate a collection of inputs with sets of (legal) collections of outputs.
A task is said to be \emph{colorless} \cite[Chap.~4.1]{HKRbook} if for any input with set of values $I$, any input formed with a subset $I'\subset I$ of values is also a possible input, and, symmetrically, for any legal output for a given $I$ with set of values $O$, an output with set of values $O'\subset O$ is also legal for $I$. 
This means its specification does not depend of the multiplicity of values in the distributed system, we only have to encode the values present in the system, rather than also keeping track of process identifiers.
So, in a colorless task specification, we must relate a complex $\In$ of input values to a complex $\Out$ of output values. Recall that a relation $R \subseteq X \times Y$ is equivalent, via \emph{currying}, to a map $f_R^X : X \fl \Pow Y$, defined by $f_R^X(x) = \{ y \in Y \mid xRy \}$. In the case of a relation between simplicial complexes, respecting the simplicial structure means that the associated map is of the form $\Delta : \In \fl \dlatt\Out$ and is required to be monotone. As recalled above in Section~\ref{SS:SimpComp}, this is known as a carrier map. In the simplicial semantics, this map is interpreted as the specification that, starting with values in $\sigma \in \In$, valid outputs are in the sub-complex $\Delta(\sigma) \in \dlatt\Out$. Such a carrier map is called a \emph{(colorless) task}. As an example, we have illustrated the \emph{binary consensus task} above on the left. This is the distributed specification in which processes start with values $0$ or $1$ and must collectively agree on one of them. The specification $\Delta$ for binary consensus is pictured in blue; when all processes have value $0$ (resp.~$1$), they must decide on that value, corresponding to the assignment $\{0\}\mapsto \{\{0\}\}$ (resp.~$\{1\}\mapsto \{\{1\}\}$). However, when at least one process has $0$ and at least on other has $1$, they can choose either value, corresponding to the assignment $\{0,1\} \mapsto \{\{0\},\{1\}\}$. Note that simplices, sets of values, are distinguished from sub-complexes, which are sets of sets of values.

\subsection{Protocol complexes and strict carrier maps}

As recalled previously, distributed computability is concerned with determining which tasks can be solved in a given communication protocol. Above we defined {colorless} tasks in the simplicial semantics. Here, we describe how {achromatic} protocols are encoded. For the definition of {colorless} task-solvability in the simplicial semantics, see Section~\ref{SS:DistTaskCat} below.

{
Note that we employ the term colorless (resp.~achromatic) to describe tasks (resp.~protocols) which can be encoded via (achromatic) simplicial complexes. As mentioned in the introduction, the precise semantic relationship between colored and colorless task-solvability with respect to chromatic or achromatic models is not well understood. For this reason we choose to terminologically distinguish tasks and protocols with respect to (achromatic) simplicial encoding. Henceforth we will consider exclusively colorless tasks and achromatic protocols, and will thus often omit the qualifications ``colorless'' and ``achromatic''.
}

Broadly speaking, a protocol consists of a set of valid communication scenarios along with their epistemic effect on the input states. Since the order of communications will affect the knowledge of each process, these protocols are inherently non-deterministic. As above, we glue these possible reachable states along common local states to obtain a simplicial complex. For example, consider a system in which at least two processes, starting with either $0$ or $1$, send each other their local value, and in which at most one message may be lost. The carrier map describing this communication protocol is pictured below on the right.
\begin{wrapfigure}{r}{.3\textwidth}
\begin{tikzpicture}[scale=.4]
      \node[] at (-1,3.5) {$\In$};
      \node[circle, fill=gray!30, draw] (0) at (0,6.0) {$0$};
      \node[circle, fill=gray!30, draw] (1) at (0,1) {$1$};
      \draw (0) -- (1);

      \node[] at (9,3.5) {$\Pc$};
      \node[circle, minimum size = 25pt, fill=gray!30, draw] (5) at (7,7.0) {$0$};
      \node[circle, minimum size = 25pt, radius =3pt, fill=gray!30, draw] (4) at (7,0) {$1$};
      \node[circle, minimum size = 6pt, fill=gray!30, draw] (6) at (7,3.5) {$01$};
      \draw (4) -- (6);
      \draw (5) -- (6);

      \node[blue] at (3,4) {$\pi$};
      
      \node[] (12) at (0,3.5) {};
      \node[] (13) at (5,3.5) {};      
      \draw[blue,label=north:${\Delta}$] (12) -- (13);
      
      \node[] (14) at (4.5,3.5) {};
      \node[] (15) at (4.8,3.1) {};
      \node[] (16) at (4.8,3.9) {};
      \node[] (17) at (7,1.75) {};
      \node[] (18) at (7,5.25) {};
      \draw[->,blue] (16) to[bend right] (4);
      \draw[->,blue] (15) to[bend left] (5);
       \draw[->,blue] (16) to[bend right] (17);
      \draw[->,blue] (15) to[bend left] (18);
      \draw[->,blue] (14) to (6);
      \draw[->,blue] (0) -- (5);
      \draw[->,blue] (1) -- (4);

\end{tikzpicture}
\end{wrapfigure}
Indeed, when all processes start with $0$ (resp.~$1$), the only value they can see is $0$ (resp.~$1$), regardless of how many messages are lost. However, when at least one process starts with $0$ and another with $1$, corresponding to the input simplex $\{0,1\}$, there are several possibilities. If no message is lost, all processes see both values, corresponding to the subcomplex $\{\{01\}\}$. If however at least one process having $0$ (resp.~$1$) has not received a message containing a different value than it already has, it sees only $0$ (resp.~1), while other processes see $01$, corresponding to the sub-complex $\{\{0\}, \{0,01\}, \{01\}\}$ (resp.~$\{\{1\},\{1,01\}, \{01\}\}$). Note that if \emph{all} messages can be lost, we must add an extra simplex to $\Pc$, namely $\{0, 1\}$ and include it in the image of $\pi(\{0,1\})$.

Summing up, an ({achromatic}) distributed protocol is given by an input complex $\In$ encoding input states, a \emph{protocol complex} $\Pc$ encoding reachable states, and a carrier map $\pi : \In \fl \dlatt\Pc$. In~\cite{HKRbook}, carrier maps associated to protocols are required to satisfy additional properties, namely
\begin{itemize}
\item $\forall \sigma_1, \sigma_2 \in \In$, $\pi(\sigma_1 \cap \sigma_2) = \pi(\sigma_1)\cap \pi(\sigma_2)$ (\emph{strictness}),
\item $\bigvee_{\sigma\in\Ic} \pi_\Ic(\sigma) = \Pc$ (\emph{effective surjectivity}).
\end{itemize} 
For this reason, we refer to carrier maps which satisfy these conditions as \emph{protocol maps}. 

The first contribution of this paper is in identifying these conditions as precisely those needed to lift a protocol map to a homomorphism of lattices from $\dlatt\In$ to $\dlatt\Pc$. Recall that given an order preserving map $f: P \fl \dlatt Q$, where $P$ and $Q$ are posets, we obtain a union-preserving map, called the \emph{join-lift} of $f$:
\begin{align*}
\jlift f : &\dlatt P \longrightarrow \dlatt Q \\
	& A \longmapsto \bigcup\{f(a) \mid a\in A\}.
\end{align*}
It is routine to check that strictness and effective surjectivity imply that the join-lift of a protocol map $\pi$ not only preserves unions, but is in fact a lattice homomorphism $\jlift\pi : \dlatt\In \fl \dlatt\Pc$. Applying Birkhoff duality, we can thus conclude that a protocol map is equivalent to a monotone map $\rho : \Pc \fl \In$. This insight naturally leads to spectral spaces as the generalization of the simplicial semantics of distributed computability, as is described in Section~\ref{S:SpecSpaces}.

The above is a very general definition of a protocol, in the sense that the protocol complex $\Pc$ can represent \emph{any} reachable configuration of distributed knowledge. In this paper, we consider \emph{round-based} protocols. This means that we have a collection of protocol complexes $(\Pc^r, \pi^r)$ for every round $r$. However, due to the fact that these protocols perform the same actions each round, in certain cases such protocols can be understood as functorial transformations of the input complex. We describe this construction in the following sub-section.

\subsection{Round-based full information protocols as endofunctors}\label{SS:ProtEndo}

For round-based, content-neutral, full-information models, the protocol complex is simply defined and appears as a functorial construction describing how the simplices (\ie global states) of the input complex are modified by the action of the protocol in one round. The full scope of the correspondence between protocols and functors is not fully known, but we describe the functorial construction for
our protocols in Appendix~\ref{A:ProtocolsAsFunctors}. More precisely, we consider protocols that are 

\begin{itemize}
    \item \emph{Round-based}: the communication scheme is temporally organized into rounds,
      the information only flows inside a given round.
    \item \emph{Full-information}: the distributed system can be
      described by the tuple of the local states {and} a
      process has no limit on the size of information it can
      share. This implies that it is possible to ``store and forward''
      all information received (doing this repeatedly is called full information
      algorithm in distributed computing).
    \item \emph{Content-neutral}: the way information is
      transmitted (or not) does not depend on the actual content of this
      information.
  \end{itemize}

For the purposes of this paper, a \emph{protocol} is a pair $(\Pi, \pi)$ where $\Pi : \scompcat \fl \scompcat$ is a functor and $\pi$ is a natural transformation between $U: \scompcat \fl \Posf$, the forgetful functor, and $V\circ\Dr\circ\Pi : \scompcat \fl \Posf$, where $V: \DLf \fl \Posf$ is the forgetful functor. Given a simplicial complex $\Cr$, we require the component $\pi_\Cr : \Cr \fl \dlatt{\Pi(\Cr)}$ to be a protocol map. {Furthermore, due to the full-information hypothesis, the join-lifts of the $\pi_\Cr$ are assumed to be injective maps, see Appendix~\ref{A:ProtocolsAsFunctors}}.
Due to this naturality condition, we may consider the components $\pi_{\Pi^n(\Cr)}$ for each $n \in \NN$, which to simplify notation below we denote by $\pi_n$, obtaining a sequence of maps in the category $\DL$:
\begin{equation}\label{E:DlattSeq}
\xymatrix{
\dlatt \Cr
\ar[r]^-{\scriptstyle\jlift{\pi}_0}
&
\dlatt{\Pi(\Cr)}
\ar[r] ^-{\scriptstyle\jlift{\pi}_1}
&
\cdots
\ar[r]^-{\scriptstyle\jlift{\pi}_2}
&
\dlatt{\Pi^n(\Cr)}
\ar[r]^-{\scriptstyle\jlift{\pi}_n}
&
\dlatt{\Pi^{n+1}(\Cr)}
\ar[r]^-{\scriptstyle\jlift{\pi}_{n+1}}
&
\cdots
}
\end{equation}
For $n\leq n'$, we denote by $\jlift{\pi}_{n,n'} : \dlatt{\Pi^n(\Cr)} \fl \dlatt{\Pi^{n'}(\Cr)}$ the composition $\jlift{\pi}_{n} \circ\cdots \circ \jlift{\pi}_{n'-1}$. We also denote by $\rho_\Cr$ the dual map associated to $\jlift{\pi}_\Cr$, thereby obtaining a natural transformation $\rho : U\circ\Pi \fl U$. This is called the \emph{dual natural transformation} associated to $\pi$. {Since duality sends monos to epis and the $\jlift\pi_\Cr$ are injective, the maps $\rho_\Cr$ are surjective}.

This functorial definition of protocol maps has also been proposed in~\cite[Section~4]{catdist-x}, but without the associated natural transformation. Moreover, it is consistent with the description of many well-studied distributed protocols. For example, the Immediate Iterated Snapshot ($IIS$) protocol corresponds functorially to the barycentric subdivision, and its carrier maps can be defined as natural transformations, see Section~\ref{SS:SpecSpaceIIS}. 

\section{Spectral semantics of distributed computability}\label{S:SpecSpaces}

The main insight of this paper is to view a simplicial complex as a finite spectral space, that is, in contrast to the usual approach, not as a combinatorial object nor as a Hausdorff topological space built from higher-dimensional triangles. Rather, we view a simplicial complex as a poset endowed with the Alexandrov topology. This establishes the appropriate relationship between simplicial complexes and their lattices of sub-complexes via Birkhoff duality, thereby also simplifying the relationship between carrier maps and simplicial maps. This construction can be taken to the (co-)limit, producing a general spectral space and a dual distributive lattice associated to a given protocol.
We fix a simplicial complex $\Ic$ for the remainder of this section. 

\subsection{From endofunctors to inverse limit systems}

Applying Birkhoff duality to the sequence (\ref{E:DlattSeq}), or equivalently using the dual natural transformation associated to $\pi$, we deduce the existence of order-preserving maps $\rho_n : \Pi^{n+1}(\Ic) \fl \Pi^n(\Ic)$ on the underlying simplicial complexes. This gives rise to a projective limit system
\begin{equation}\label{E:InvLim}
\xymatrix@C=3em{
\Ic
&
\Pi(\Ic)
	\ar[l]_-{\scriptstyle\rho_0}
&
\Pi^2(\Ic)
	\ar[l]_-{\scriptstyle\rho_1}
&
\cdots
\ar[l]_{\scriptstyle\rho_2}
&
\Pi^{n-1}(\Ic)
	\ar[l]_-{\scriptstyle\rho_{n-2}}
&
\Pi^{n}(\Ic)
	\ar[l]_-{\scriptstyle\rho_{n-1}}
&
\cdots
	\ar[l]_-{\scriptstyle\rho_{n}}
}
\end{equation}
in the category of spectral spaces, since each $\Pi^n(\Ic)$ is a spectral space when endowed with the Alexandrov topology.
The theory of spectral spaces~\cite{Hoc69} tells us that the limit of this diagram, which we denote by $\speclim \Ic$, is a spectral space. We denote by $\rho_{n',n} :  \Pi^{n'} \fl \Pi^{n}$ the composition $\rho_{n'} \circ \cdots \circ \rho_{n-1}$, \ie the dual of $\jlift\pi_{n,n'}$, and by $\lambda_n : \speclim \Ic \fl \Pi^n(\Ic)$ the canonical projections associated with the limit construction. Note that since each $\rho_n$ is surjective, so are the $\lambda_n$.

Moreover, since we consider up-sets as open sets, the colimit of the dual diagram (\ref{E:DlattSeq}) which we denote by $\dlattlim \Ic$, is the lattice of closed sets of $\speclim\Ic$ whose complements are compact, which generate its closed sets under arbitrary intersections and finite unions. We denote by $i_n : \dlatt{\Pi^n(\Ic)} \fl \dlattlim\Ic$ the canonical maps associated with the colimit, {which by duality are injective, see Remark~\ref{R:OrdDual}}. 
Summing this up, we obtain our first theorem.
\begin{theorem}
A protocol $(\Pi, \pi)$ defines functors 
$
\specinfty : \scompcat \fl \Spec
\text{ and }
\dinfty : \op\scompcat \fl \DL,
$
such that for any simplicial complex $\Ic$, $\dlattlim \Ic$ generates the closed sets of $\speclim \Ic$.
\end{theorem}
\begin{proof}
The assignments $\Ic \mapsto \dlattlim \Ic, \speclim \Ic$ are well defined by co-completeness of $\DL$ and since directed limits of finite $T_0$ spaces are spectral (see \cite{Hoc69}, Proposition 10), respectively. Functoriality is a consequence of the universal property of (co-)limits. Explicitly, for a simplicial map $\phi : \Ic \fl \Ic'$, we deduce morphisms $\Pi^n \phi: \Pi^n(\Ic) \fl\Pi^n(\Ic')$ meaning that $\Pi^\infty(\Ic)$ is a cone under $(\Ic' \leftarrow \Pi(\Ic') \leftarrow \cdots)$. By the universal property of limits, we obtain a unique arrow
$
\Pi^\infty \phi : \Pi^\infty(\Ic) \fl \Pi^\infty(\Ic').
$
For $\dinfty$ the proof is similar except for that the dual universal property for colimits implies contravariance.

Under Stone duality, the dual lattice to $\speclim \Ic$ is its set of compact-opens. In terms of colimits, this means we are considering the spectral space dual to the colimit $\ulattlim \Ic$ of the sequence 
\begin{equation}\label{E:FiltColimit}
\xymatrix{
\ulatt \Ic
\ar[r]%
&
\ulatt{\Pi(\Ic)}
\ar[r] %
&
\cdots
\ar[r]%
&
\ulatt{\Pi^n(\Ic)}
\ar[r]%
&
\ulatt{\Pi^{n+1}(\Ic)}
\ar[r]%
&
\cdots
}
\end{equation}
where $\ulatt{\Pi^n(\Ic)}$is the up-set lattice of $\Pi^n(\Ic)$ and the maps are those induced from (\ref{E:InvLim}) by Birkhoff duality. Since each $\dlatt{\Pi^n(\Ic)}$  consists of the complements of $\ulatt{\Pi^n(\Ic)}$, we see that this means that $\dlattlim \Ic$ is the set of closed sets in $\speclim \Ic$ whose complements are compact-open, and thus generates its closed sets.
\end{proof}

\begin{remark}\label{R:OrdDual}
We underline here that, following the discussion at the end of the above proof, the maps $i_n$ associated with the colimit construction are not the dual maps of the $\lambda_n$. Indeed, the dual maps of the $\lambda_n$ are maps $\ulatt{\Pi^n(\Ic)} \fl \copen{\speclim\Ic}$, where $\ulatt{\Pi^n(\Ic)}$ and  $\copen{\speclim\Ic}$ are the order duals of  $\dlatt{\Pi^n(\Ic)}$ and $\dlattlim\Ic$, respectively. The sequence above involving up-set lattices is obtained from (\ref{E:InvLim}) by Birkhoff duality, so its maps are the order duals of the $\jlift{\pi_n}$, meaning that the $i_n$ are the order duals of the dual maps of the $\lambda_n$.
\end{remark}

\subsection{The spectral space of a protocol}

Here we describe the spectral space $\speclim \Ic$ concretely. Indeed, we show that points in $\speclim \Ic$ correspond to certain sequences of simplices and explicitly describe its topology. This will lead us nicely to our main result, namely that the topology on $\speclim \Ic$ characterizes task solvability, see Theorem~\ref{thm-carac} below. 
To this end, we consider sequences $(\sigma_n)_n$ of simplices such that for all $n$,
\begin{itemize}
\item $\sigma_n \in \Pi^n(\Ic)$,
\item $\sigma_{n+1} \in \pi_n(\below \sigma_n)$, or equivalently, $\rho_n(\sigma_{n+1}) = \sigma_n$.
\end{itemize}
We call these \emph{protocol sequences}.
Since $\speclim \Ic$ is the limit of the sequence (\ref{E:InvLim}), we have \cite[Section~2.3.9]{spectralBook} that the underlying set of $\speclim \Ic$ is 
$
\{(\sigma) \in \prod \Pi^n(\Ic) \mid \rho_n(\sigma_{n+1}) = \sigma_n \},
$
that is, the set of protocol sequences.  
By \cite{spectralBook}, the specialization order $\preceq$ on $\speclim\Ic$ is determined point-wise. More explicitly, given protocol sequences $(\sigma)$ and $(\tau)$, we have $(\sigma)\preceq(\tau)$ if, and only if, for all $n$ $\sigma_n \preceq_n \tau_n$, where $\preceq_n$ is the specialization order in the finite $T_0$ space $\Pi^n(\Ic)$. The latter is given by $
\sigma_n \preceq_n \tau_n
\iff
\above\sigma_n \subseteq \above\tau_n,
$
whereby we deduce $(\sigma)\preceq(\tau)$ if, and only if, $\sigma_n \geq \tau_n$ for all $n$.
Finally, we have the following characterization of $\speclim\Ic$. See also Appendix~\ref{SS:SpecSpaceIIS} for a full description of the spectral space associated to the $IIS$ protocol.

\begin{proposition}\label{P:ProtSpecSpace}
Given a protocol $(\Pi, \pi)$, we obtain a functor $\specinfty : \scompcat \fl \Spec$ sending a simplicial complex \In to the spectral space $\speclim\Ic$, such that
\begin{itemize}
\item The points of $\speclim\Ic$ are in one-to-one correspondence with protocol sequences $(\sigma)$,
\item The sets $U_{n,v} := \{ (\tau) \in \speclim\Ic \mid \tau_n \geq \sigma\} = \bck \lambda_n(\above v)$ form a sub-basis for the open topology.
\item The specialization order $\preceq$ on $\Pi^\infty(\Ic)$ is defined by
$
x_\sigma \preceq x_\tau
\iff
\forall n, \sigma_n \geq \tau_n,
$
\ie the opposite order of the point-wise order on protocol sequences.
\end{itemize}
\end{proposition}

\section{Colorless Computability}\label{S:ColorlessComp}

In this section, we prove our main result, namely a topological characterization of colorless computability. 
First, we recall the formal definition of solving a task from~\cite{HKRbook} and reformulate via Stone duality to be better adapted to our proof methods.

\subsection{Generalized computability}
\label{SS:DistTaskCat}

We say that a protocol \emph{solves} a task $(\Ic, \Oc, \Delta)$ if there exists an $n \in \NN$ and a simplicial map $\delta : \Pi^n(\Ic) \fl \Oc$, such that for all $\sigma\in\Pi^n(\Ic)$, $\fwd\delta(\pi_{0,n}(\sigma))\subseteq\Delta(\sigma)$.
Remarking that in full information models, distributed algorithms have a canonical presentation as \emph{full information algorithms}, 
this is the classical definition of the simplicial semantics of distributed computing
(see \eg \cite{HKRbook}), which we reformulate exclusively in terms of down-set lattices. 
Unwinding the definitions of the join-lift of a map and using the adjunction between inverse and direct images, one easily proves that the diagram on the left sub-commutes if, and only if, the diagram on the right sub-commutes, \ie for $A \in \dlatt\Ic$, $\jlift\pi_{0,n}[A] \leq \bck\delta[\jlift\Delta [A]$.
Note that the right-hand diagram is not in $\DL$ since $\jlift\Delta$ only preserves joins in general.
\begin{equation}\label{E:TaskSolve}
\xymatrix{
 \Ic
	\ar[rr] ^{\Delta}
	\ar[dr] _{{\pi}_{0,n}}
&
&
\dlatt \Oc
\\
&
\dlatt{\Pi^n(\Ic)}
	\ar[ur] _{\fwd \delta}
	\ar@{}[u] |-{\mathbin{\rotatebox[origin=c]{90}{$\leq$}}}
&
}
\qquad\qquad
\xymatrix{
\dlatt \Ic
	\ar[rr] ^{\jlift\Delta}
	\ar[dr] _{\jlift{\pi}_{0,n}}
&
&
\dlatt \Oc
	\ar[dl] ^{\bck \delta}
\\
&
\dlatt{\Pi^n(\Ic)}
\ar@{}[u] |-{\mathbin{\rotatebox[origin=c]{45}{$\leq$}}}
&
}
\end{equation}

Given a task $(\Ic, \Oc, \Delta)$ and a protocol $(\Pi, \pi)$, we say that a spectral map $f : \speclim \Ic \fl \Oc$ is \emph{carried} by $\Delta$ provided that $i_0[A] \leq \bck f[\jlift\Delta[A]]$.

Finally, we recall that the \emph{geometric realization} $\real\Cr$ of a simplicial complex $\Cr$ is, roughly speaking, the interpretation of each $n$-simplex as the convex hull of $n+1$ linearly independent points in $\RR^{n+1}$, glued along common sub-simplices, see~\cite[p.~7]{goerss1999simplicial} for a more detailed account. As above for spectral maps, we say that a continuous map $f: \real\In \fl \real \Oc$ is \emph{carried} by $\Delta$ provided that, for all $\sigma \in \In$, $f(\real\sigma) \subseteq \real{\Delta(\sigma)}$. In~\cite[Thm. 4.2.3]{HKRbook}, the authors show that for the $IIS$ protocol, task-solvability is equivalent to the existence of a continuous map $\real\In\fl\real\Out$ carried by $\Delta$, see Theorem~\ref{Theorem1} in the Appendix. The goal of this section is to generalize this groundbreaking topological characterization of task-solvability.

\subsection{Distributed Computability theorem}
Before formulating our main result, we first prove the following lemma:
\begin{lemma}\label{L:SpecFact}
Given a a spectral map $f : \speclim \Ic \fl \Oc$, there exists $n$ such that $f$ {sub-}factorizes as $\speclim \Ic \overset{\lambda_n}{\longrightarrow} \Pi^n(\Ic) \overset{\delta}{\longrightarrow} \Oc$, \ie $\delta\circ\lambda_n \leq f$, where $\delta$ is a simplicial map, and $\lambda_n$ is the canonical projection.
\end{lemma}
\begin{proof}

The family $(\above w)_{w \in V_\Oc}$ is an open covering of $\Oc$, so $(\bck f(\above w))_{w \in V_\Oc}$ is a covering of $\speclim \Ic$. Since $f$ is spectral, each of the sets $\bck f(\above w)$ is compact-open in $\speclim\Ic$, so there exist finite families $(n^w_i)$ and $(\sigma^w_i)$ so that $\bck f(\above w) = \bigcup_i \bck\lambda_{n^w_i}(\above \sigma^w_i)$ and $\sigma^w_i \in \Pi^{n^w_i}(\Ic)$. Since (\ref{E:FiltColimit}) is a filtered colimit {with injective maps}, we can choose a uniform $n$, so without loss of generality, there exists a finite family $(\sigma^w_j) \subseteq \Pi^n(\Ic)$ with $\bck f(\above w) = \bigcup_j \bck\lambda_n(\above \sigma^w_j)$. Putting this together, we conclude that the finite union $\bigcup_{w\in V_\Oc, j} \bck\lambda_n(\above \sigma^w_j)$ is an open cover of $\speclim\Ic$.

This implies that for every $v \in V_{\Pi^n(\Ic)}$, there exists a $w_v \in V_\Oc$ so that $\bck\lambda_n(\above v) \subseteq \bck f(\above w_v)$. Indeed, since $\lambda_n$ is surjective, there exists some $x \in \speclim\Ic$ with $\lambda_n(x) = v$, and since the family $ \bck\lambda_n(\above \sigma^w_j)$ covers $\speclim\Ic$, there exists $\sigma^w_k$ with $x \in \bck\lambda_n(\above \sigma^w_k)$ and so $v = \lambda_n(x) \geq \sigma^w_k$. By minimality of $v$, we conclude $v = \sigma^w_k$. In particular, since $f(\above w) = \bigcup_j \bck\lambda_n(\above \sigma^w_j)$, we have $\bck\lambda_n(\above v) = \bck\lambda_n(\above \sigma^w_k)  \subseteq \bck f(\above w)$. We denote this $w \in V_\Oc$ by $w_v$.

Define a map $d: V_{\Pi^n(\Ic)} \fl V_\Oc$ by $d(v) = w_v$. We verify that $\delta = \fwd d$ is simplicial. To that end, let $\sigma = \{v_1, \dots, v_t \}$ be a simplex of $\Pi^n(\Ic)$. Since $\lambda_n$ is surjective, $f(\bck{\lambda_n}(\above \sigma))$ is non-empty, so we can choose $\sigma'\in f(\bck{\lambda_n}(\above \sigma))$. Furthermore, we have 
$
f(\bck{\lambda_n}(\above \sigma)) \subseteq f(\bck{\lambda_n}(\above v_i)) \subseteq \above d (v_i)$
for all $i$ since $\bck{\lambda}_n(\above v_i) \subseteq \bck f(\above w_{v_i})$. In particular, $\sigma' \in \bigcap_i \above d(v_i)$, \ie $\delta(\sigma) = \{d(v_1), \dots, d(v_t)\} \subseteq \sigma'$, from which we conclude that $\delta(\sigma)$ is a simplex in $\Oc$.
Finally, we have that $\delta\circ\lambda_n \leq f$. Indeed, for $v \in \lambda_n(x)$, we have $\bck{\lambda_n}(\above v) \subseteq \bck f(\above d(v))$ so in particular $f(x) \in \above d(v)$, \ie $d(v) \in f(x)$ for each $v \in \lambda_n(x)$, showing the sub-factorization.
\end{proof}

\begin{remark}
  The above proof is similar to the proof using simplicial approximation found in~\cite[Proposition 3.7.3]{HKRbook}, generalized to the case of spectral maps. Indeed, the conclusion is reached by showing that $f$ satisfies the so-called \emph{star condition}, namely that there exists $n$ such that for all $v\in V_{\Pi^n(\Ic)}$, there exists $w \in V_\Oc$ such that $f(\bck{\lambda}_n(\above v)) \subseteq \above w$.
\end{remark}

\begin{theorem}\label{thm-carac}
A protocol $(\Pi, \pi)$ solves a colorless distributed task $\Delta: \Ic \fl \dlatt \Oc$ if, and only if, there exists a spectral map $f:\speclim \Ic \fl \Oc$ carried by $\Delta$.
\end{theorem}
\begin{proof}
If $(\Pi, \pi)$ solves the task, there exists $n$ and a simplicial map $\delta : \Pi^n(\Ic)\fl \Oc$ so that $\jlift\pi_{0,n} \leq \bck\delta \circ \jlift\Delta$. Then $f = \delta \circ \lambda_n$ is a spectral map which is carried by $\Delta$. Indeed, we have $i_0 = i_n \circ \jlift\pi_{0,n} \leq i_n \circ \bck\delta \circ \jlift\Delta = \bck f \circ \jlift\Delta$. The final equality holds since $\bck f : \dlatt \Oc \fl \dlattlim\Ic$ is the order dual of $\copen{f}$ and $i_n$ is the order dual of $\copen{\lambda_n}$, see Remark~\ref{R:OrdDual}.

Conversely, given a spectral map $f:\speclim \Ic \fl \Oc$ and applying Lemma~\ref{L:SpecFact}, we obtain a simplicial map $\delta : \Pi^n(\Ic) \fl \Oc$ satisfying $\delta\circ\lambda_n \leq f$. Routine verification shows that this implies that $\copen{\lambda_n}\circ\copen{\delta} \leq \copen{f}$, so by order duality we have $\bck f \leq i_n \circ \bck \delta$. By hypothesis, $i_0 \leq \bck f\circ \jlift\Delta$, so substituting $i_0 = i_n \circ \jlift\pi_{0,n}$ yields $i_n \circ \jlift\pi_{0,n} \leq \bck f\circ \jlift\Delta = i_n \circ \bck \delta\circ \jlift\Delta$. By injectivity of $i_n$, we obtain $\jlift\pi_{0,n} \leq \bck\delta\circ\jlift\Delta$, \ie $\Delta$ is carried by $\delta$.
\end{proof}

\begin{remark}
When the protocol is Iterated Immediate Snapshot $(IIS$), this computability theorem is exactly the layered (round-based) version of the celebrated Asynchronous Computability Theorem (ACT) of Herlihy and Shavit \cite{HS99}. This is further elaborated in Appendix~\ref{S:IIS}.
\end{remark}

\section{Conclusion}

In this paper we have introduced a novel topological encoding for
round-based, content-neutral, full-information achromatic distributed protocols via spectral spaces
and shown that this leads to a characterization of colorless computability generalizing the classical colorless computability theorems.
These preliminary results demonstrate the effectiveness of this approach.
However, we envisage many extensions of the present state of the spectral semantics. First, we position our work with respect to other topological approaches to task-solvability.

\paragraph*{Comparison with previously defined spaces}
\label{sec-prevsp}
Before comparing our contribution with previously defined spaces,
we summarize %
previous works related to topological methods in distributed computing.  Topological methods were pioneered by Herlihy and
Shavit~\cite{HS99}, Saks and Zaharoglou \cite{SZ}. This yielded a first wave of applications that
were collated in the book of Herlihy, Koslov and Rajsbaum
~\cite{HKRbook}.
A second wave is currently extending the results to many other models
than the asynchronous wait-free read-write, $t-$resilient
or $IIS$ models that are considered in \cite{HKRbook}.
Recently, there have been
two lines of work toward a generalization. In
\cite{consensus-epistemo,ACN23}, arbitrary round-based models \ma
were considered and it was proved that there exists a general
topological space $\mathcal E_\ma$ (first introduced for the
Consensus task in \cite{consensus-epistemo}) such that a colored task
$(\In,\Out,\Delta)$ is solvable against \ma if and only if there is a
chromatic simplicial map $f:\mathcal
E_\ma\fl\Out$~\cite[Thm.5.4]{ACN23}. 
In \cite{2generals-journal,CG-geoconf,CG-24}, the generalization is furthered by
considering arbitrary subsets of executions of $IIS$. Many classical asynchronous
models can be presented this way, so this was a step towards a general
computability theorem. However the results that were obtained in
\cite{CG-24} are only for colorless tasks : a colorless task
$(\In,\Out,\Delta)$ is solvable in model $\ma\subset IIS$ if and 
only if there is a continuous function $f:geo(\In\times\ma)\fl\Out$,
where $geo$, the ``geometrization mapping'', sends a model to a subset
of $|\In|$. %

The first line of work is more general (more adversaries, all tasks)
but has somewhat abstract topological statements, while the second line of
work has simpler geometric statements but only for specific adversaries
and colorless tasks. To see why, despite the generality of the
approach, it is not straightforward to derive the results from the
second line of work from the first, one could compare the
statement for computability of set-agreement in sub-$IIS$ models in
\cite[Thm.~4.2]{ACN23} and \cite[Thm.~25]{CG-geoconf,CG-24}.  By showing that a general topological
representation of a distributed system should be a spectral space, and leveraging the Stone duality,
our contribution unifies both approaches and takes the best of
both lines of work~: general topological statements for distributed
computability that are amenable to simple geometric proofs for
colorless tasks.

\medskip
We underline that our spectral space $\Pi^\infty_\ma(\In)$ is not the
space from \cite[Theorem 5.4]{ACN23}, which is Hausdorff; nor from
\cite{consensus-epistemo}, which is chromatic, but also contains only executions.
Our approach using spectral spaces allows us to separate distinct executions while retaining a strong topological relation when they are indistinguishable in the sense of distributed computation. This is now apparent within the specialization order of the spectral topology, since executions are indistinguishable precisely when they are below the same maximal point w.r.t.~the specialization order.
In \cite{CG-geoconf}, the chromatic IIS model is
investigated, that is, the standard chromatic subdivision protocol.
The set of executions is projected onto a geometric
subset of $\RR^N$ by a $geo$ mapping. In \cite[Th.~25]{CG-geoconf}, it is shown that it
is possible to classify the geometric points by the number of
pre-images in the set of executions~: 1, 2 or infinity. This would have
provided a similar classification if the achromatic $IIS$ model had been considered. 
This agrees with our
Prop.~\ref{P:CherryCount} classifying down-sets in the specialization order. 
 As noted previously, execution sequences,
\ie protocol sequences consisting of maximal simplices, are in one-to-one
correspondence with executions. So it seems
that the spectral space contains both the geometric points and the
associated executions, with the following structure~: whenever a point
has only one execution as pre-image, it corresponds to an execution sequence,
whereas, when the point has at least two pre-images, we have both the
geometric point, as a maximal point in the spectral space, and the associated execution sequences as minimal points in its down-set under the specialization order. So the spectral space we construct could be interpreted as containing the merge (without unnecessary duplication) of the set of execution sequences and the set of ``limit geometrization'' points, as well as all intermediary protocol sequences, all organized into a nice $T_0$ space. This is made explicit for the achromatic $IIS$ protocol in Appendix~\ref{S:IIS}.

\paragraph*{Future Work}
\label{sec:ccl}
Future extensions include determining equivalence of models by studying the topological relationship between the associated spectral spaces, {as indicated by Theorem~\ref{Cor:ChComp}}, but also investigating how to consider their subspaces as submodels in order to treat so called non-compact models.
Since submodels of $IIS$ have been proved \cite{CG-24} to have the same kind of geometric computability characterization, we believe this should be achievable in the framework of this paper.

It is also important
to determine how to describe general (colored) tasks using a
similar spectral approach. {As briefly discussed in the introduction, this requires a mathematical investigation of projective limits of chromatic simplicial complexes, and in particular defining a notion of chromatic spectral space. In addition, in order to obtain a unified framework, a minute understanding of the semantic relationship between chromatic and achromatic models, as well as between colored and uncolored tasks, must be established. These two directions constitute immediate future work, which will be informed and ramified by the description of chromatic models in related works. In particular, we believe that the algebraic approach given in~\cite{catdist-x} is highly related to our topological approach via Stone duality; the dual of the appropriate chromatic spectral space should precisely correspond to the limit algebra they define. Furthermore, the topological approach to chromatic models given in~\cite{ACN23} is given by an ultra-metric, these being well-known to correspond to Priestley spaces, which are derivable from spectral spaces via the patch topology construction.} 

{Finally, an extension to the case of asynchronous protocols is necessary to achieve a unified semantics. As opposed to round-based protocols, processes are allowed to continue communicating even if some processes are very slow. Describing such protocols also constitutes a future research direction.}
\def\cprime{$'$}

\appendix

\section{Applications to IIS model}\label{S:IIS}

Here we describe the applications of spectral topology and duality theory to a specific model of distributed computing. This culminates in proving, via novel methods, the following classical result:

\begin{theorem}[Colorless Distributed Computability Theorem for achromatic IIS]\cite[Thm.~4.2.3]{HKRbook}\label{Theorem1}
Let $(\Ic,\Oc,\Delta)$ be a colorless task. It is solvable by an Iterated Immediate Snapshot protocol if and only if there
exists a continuous map $f: \real{{\Ic}} \fl \real\Oc$
carried by $\Delta$.
\end{theorem}

 \begin{theorem}[Colorless Distributed Computability Theorem for (chromatic) IIS]\cite[Thm.~4.3.1]{HKRbook}\label{Theorem2}
 Let $(\Ic,\Oc,\Delta)$ be a colorless task {for a chromatic protocol on $n+1$ processes}. It is solvable by a
 chromatic immediate iterated snapshot protocol for $n+1$ processes if and only if there
 exists a continuous map $f: \real{{\Ic}} \fl \real\Oc$
 carried by $\Delta$.
 \end{theorem}

 \begin{theorem}[Strong Colorless Distributed Computability Theorem for IIS]\label{Theorem3}
 Let $(\Ic,\Oc,\Delta)$ be a colorless task.  It is solvable by the achromatic immediate iterated snapshot protocol if and only if it is solvable by the chromatic immediate iterated snapshot protocol {for any number of processes}.
 \end{theorem}

 \noindent Theorem~\ref{Theorem3} is a consequence of the previous two. 
In this section, we show that mesh-shrinking subdivision protocols, chromatic or achromatic, all have equivalent colorless task-solvability, thereby directly obtaining the above result by more general arguments. We also prove Theorem~\ref{Cor:ChComp}, which allows us to recover Theorem~\ref{Theorem3} via natural transformations between the functors associated to the achromatic and chromatic $IIS$ protocols. We then finish by providing a full description of the spectral spaces associated to the achromatic $IIS$ protocol.
However, first we need to define colorless task-solvability for chromatic protocols.

 \subsection{Chromatic protocols}

 Recall that an \emph{$n$-labeling} of a simplicial complex $\Cr$ is a map $l:V_\Cr \fl \{1,\dots,n\}$. A \emph{chromatic simplicial complex} is an $n$-labeled simplicial complex such that for all $\sigma\in \Cr$, the restriction $l_{|\sigma} : \sigma \fl \{1,\dots,n\}$ is injective. Given chromatic complexes $(\Cr, l)$ and $(\Cr', l')$, a simplicial map $f:\Cr \fl \Cr'$ is \emph{chromatic} if for every $v\in V_\Cr$, $l(v) = l'(f(v))$. Similarly, a carrier map $\phi : \Cr \fl \dlatt{\Cr'}$ is chromatic if $\phi$ is rigid and for all $\sigma \in \Cr$, we have $l(\sigma) = l'[\phi(\sigma)]$. The category of chromatic simplicial complexes and chromatic simplicial maps is denoted by $\chrcompcat$. Finally, we recall that a simplicial complex is \emph{pure} if every maximal simplex $\sigma\in\Cr$ satisfies $\dim(\sigma) = \dim(\Cr)$.

 In~\cite{HKRbook}, a protocol for $(n+1)$ processes is defined as a triple $(\In, \Pc, \pi)$ where $\In$ and $\Pc$ are pure, $n$-dimensional chromatic complexes, and $\pi : \In \fl \dlatt\Pc$ is a chromatic carrier map. Adapting this to our functorial approach, we will consider a \emph{chromatic protocol} to be a pair $(\Pi, \pi)$ where $\Pi$ is an endofunctor on $\chrcompcat$ and $\pi : U' \Rightarrow V\circ\Dr\circ U\circ \Pi$, where $U' : \chrcompcat \fl \Posf$ is the forgetful functor, such that every component $\pi_\Cr$ is a chromatic carrier map. 
As in the case of colorless protocols, for each chromatic input complex $\In$, this data gives a projective limit system in the category $\Posf$, meaning we obtain a functor $\specinfty : \chrcompcat \fl \Spec$. 

 A \emph{colorless task} for a chromatic protocol on $n+1$ processes is a triple $(\In, \Out, \Delta)$, where $\In$ is a pure, $n$-dimensional chromatic complex, $\Out$ is a simplicial complex, and $\Delta$ is a carrier map. A chromatic protocol \emph{solves} a colorless task provided that there exists $n \in \NN$ and a simplicial map $\delta : \Pi^n(\Ic) \fl \Out$ such that the diagrams in $(\ref{E:TaskSolve})$ sub-commute. Note that neither $\Delta$ nor $\delta$ are chromatic in general. All of our constructions for achromatic protocols extend to colorless task-solvability for chromatic protocols, meaning that Theorem~\ref{thm-carac} equally applies to chromatic protocols and their colorless tasks.

\subsection{Mesh-shrinking subdivision operators}

Here we refine our characterization of spectral spaces obtained from protocols in the case of mesh-shrinking protocols, \ie those for which the associated endofunctor on simplicial complexes $\Pi$ is a mesh-shrinking subdivision operator. Indeed, in this case, the spectral space contains the geometric realization of the considered input complex $\In$ as its subspace of maximal points.

A \emph{mesh-shrinking subdivision operator} is a functor $\Pi: \scompcat \fl \scompcat$ such that 
\begin{itemize}
\item $\real{\Pi(\Cr)}$ is homeomorphic to $\real\Cr$ (\emph{subdivision}),
\item The diameter of realizations of protocol sequences vanish (\emph{mesh-shrinking}), that is, $diam(\real{\sigma_n})\overset{n\fl\infty}{\longrightarrow}0$ 
where, for $A \subseteq \real\Cr$, $diam(A) = \max\{d(x,y) \mid x,y \in A \}$.
\end{itemize}

Now, consider a protocol $(\Pi, \pi)$ where $\Pi$ is a mesh-shrinking subdivision operator. We write $R_n = \{ \real A \mid A \in \dlatt{\Pi^n(\Ic)} \}$ for the lattice of realizations of elements of $\dlatt{\Pi^n(\Ic)}$. Since $\real{A\cup B} = \real A \cup \real B$ and $\real{A\cap B} = \real A \cap \real B$, we have $R_n \cong \dlatt{\Pi^n(\Ic)}$. Abusing notation, we will write $\pi_n : R_n \fl R_{n+1}$. The colimit of these maps gives a basis for the closed sets of $\real \In$. Before showing this, we need some definitions.
Given a point $x \in \real \In$, let
\begin{itemize}
\item $\sigma_n^x$ the unique simplex of minimal dimension in $\Pi^n(\In)$ such that $x \in \real{\sigma_n^x}$,
\item $C_n^x := \below(\above\sigma_n^x)$ the closed star of $\sigma_n^x$.
\item $\Gamma_n^x := \{\sigma \in \Pi^n(\In) \mid \sigma_n^x \not\leq \sigma\}$,
\end{itemize}
By construction, $\real{\sigma_n^x}$ is the smallest closed set containing $x$ in $R_n$, and we have $\real{\sigma_{n+1}^x} \subseteq  \real{\sigma_{n}^x}$, so $\bigcap \real{\sigma_n^x} = \{x\}$. 
Moreover, observing that the set $\Gamma_n^x$ is a downset in $\Pi^n(\In)$, and the complement of its realization is $\interior{\real{C_n^x}}$, the interior of $\real{C^x_n}$, we conclude that the latter is the smallest open containing $x$ which is obtainable from complementing elements of $R_n$.
Finally, since $\Pi$ is mesh-shrinking, we know that the diameters of the sequences $(\real{\sigma_n^x})_n$ and $(\real{C_n^x})_n$ go to zero as $n$ goes to infinity.

\begin{lemma}\label{L:GenRealTop}
The sets $(\interior{\real{C_n^x}})$ for $x\in \real\Ic$ and $n\in \NN$ generate the open topology of $\real\Ic$. The colimit $R$ of the diagram of maps $(\pi_n : R_n \fl R_{n+1})$ is a basis for its closed sets.
\end{lemma}
\begin{proof}
Let $x \in \real \In$ and $X$ be a closed set in $\real \In$ not containing $x$. By the above remarks, we know that there exists $n$ such that $\interior{\real{C_n^x}} \subseteq \compl{X}$. Thus $X \subseteq \real{\Gamma_n^x}$, and by construction $x \not\in \real{\Gamma_n^x}$.
\end{proof}

This result states that the co-frame of closed sets for $\speclim \Ic$ and $\real \Ic$ are isomorphic. This is sufficient to conclude that $\real \Ic$ is homeomorphic to the subspace of maximal points of $\speclim\Ic$ by the works pioneered by Wallman and Frink, see Cornish~\cite{Corn74} for more details.

\begin{proposition}\label{P:MaxPoints}
Given a mesh-shrinking protocol $(\Pi, \pi)$ and a simplicial complex $\In$, the geometric realization $\real \In$ is isomorphic to the subspace of $\Pi^\infty(\In)$ consisting of maximal points with respect to the specialization order, and the map $\max : \speclim\Ic \fl \real\Ic$ is continuous.

Moreover, each point $x\in \real \Ic$ corresponds to the protocol sequence $(\sigma_n^x)$, and its principal down-set in the specialization order is order-dual to the set of protocol sequences $(\sigma)$ such that $x\in \bigcap \real{\sigma_n}$, ordered point-wise.
\end{proposition}

\subsection{Final proofs}
Theorems~\ref{Theorem1} and~\ref{Theorem2}
are direct consequences of Theorem~\ref{thm-carac} and the following lemma.
First we need the following definition: given a task $(\Ic, \Oc, \Delta)$ and a continuous map $\real \Ic \fl \real \Oc$, we say that $f$ is \emph{carried} by $\Delta$ provided that for every $\sigma\in\Ic$, $f(\real{\sigma}) \subseteq \real{\Delta(\sigma)}$.
\begin{lemma} \label{L:equivShrink}
For a mesh-shrinking protocol $(\Pi, \pi)$, and a task $(\Ic, \Oc, \Delta)$, there exists a continuous map $\real \Ic \fl \real \Oc$ carried by $\Delta$ if, and only if, there exists a spectral map $\speclim \Ic \fl \Oc$ carried by $\Delta$.
\end{lemma}
\begin{proof}
Suppose there exists a continuous map $f : \real\Ic \fl \real\Oc$ carried by $\Delta$. Note that $\real{\Pi^n(\Ic)} \cong \real\Ic$ for all $n$, so we can consider this to be the domain of $f$. Now, take a vertex $w\in \Oc$ and consider $\real{\above w}$. Since $\above w$ is the complement of a down-set, its realization is open. Similarly, for any vertex $v \in \Pi^n(\Ic)$, $\real{\above v}$ is open, and since $\Pi$ is mesh-shrinking, these opens become arbitrarily small for large enough $n$. Thus, for large enough $n$, for every vertex $v\in \Pi^n(\Ic)$, there exists some vertex $w\in \Oc$ with $\real{\above v} \subseteq\bck f[\real{\above w}]$. We write $d(v) = w$, and show that $\delta = \fwd d$ is simplicial. Let $\sigma = \{v_0, \dots, v_k\} \in \Pi^n(\Ic)$ be a $k$-simplex. We have $\interior{\real\sigma} \subseteq \real{\above v_i}$ for all $i$, meaning in particular that $f(\interior{\real\sigma}) \subseteq \bigcap \real{\above d(v_i)}$. This intersection is non-empty if, and only if, there is a simplex $\tau$ in $\Oc$ containing the vertices $d(v_i)$, meaning that $\delta(\sigma)$ is indeed a simplex in $\Oc$. Thus we have a simplicial map $\delta : \Pi^n(\Ic) \fl \Oc$ which is easily verified to be carried by $\Delta$. Precomposing by the projection $\lambda_n : \speclim\Ic \fl \Pi^n(\Ic)$ gives the desired spectral map.

Conversely, suppose we have a spectral map $g: \speclim\Ic \fl \Oc$ carried by $\Delta$. By Lemma~\ref{L:SpecFact}, we obtain an order-preserving map $\delta : \Pi^n(\Ic) \fl \Oc$ carried by $\Delta$, where we can take $n$ to be arbitrarily large. Since $\Pi$ is mesh-shrinking, for large enough $n$ and $v$ a vertex of $\Pi^n(\Ic)$, we have $\real{\fwd\delta[\above v]} \subseteq \real{\above w}$ for some vertex $w$ of $\Oc$. This is enough to conclude that the map $\delta$ is simplicial, and therefore that its realization is continuous. Since $\real{\Pi^n(\Ic)} \cong \real\Ic$, we obtain the desired continuous map $\real\Ic \fl \real\Oc$.
\end{proof}

Now we provide a new result for comparing the colorless task-solvability of a chromatic or achromatic protocol with that of the achromatic $IIS$ using natural transformations.

\begin{theorem}\label{Cor:ChComp}
Let $(\Pi, \pi)$ be an achromatic protocol, and denote by $(\Ch, b)$ the $IIS$ protocol and $\rho$ (resp.~$r$) the dual natural transformation associated to $\pi$ resp.~$b$). If there exists a natural transformation $\alpha : U\circ\Pi \Rightarrow \Ch$ such that $r_\Cr \circ \alpha_\Cr = \rho_\Cr$ for all complexes $\Cr$, then any task solvable by the $IIS$ protocol is also solvable by $\Pi$. If in addition there exists a natural transformation $\beta : \Ch_{|\scompcat} \Rightarrow \Pi$, then $\Pi$ can solve precisely the same tasks as $\Ch$.

If $(\Pi,\pi)$ is a chromatic protocol, and there exists a natural transformation $\alpha : U'\circ\Pi$ such that $r_\Cr \circ \alpha_\Cr = \rho_\Cr$ for all complexes $\Cr$, then any task solvable by the achromatic $IIS$ protocol is also solvable by $\Pi$. If in addition there exists a natural transformation $\beta : \Ch\circ U'' \Rightarrow U''\circ\Pi$, then $\Pi$ can solve precisely the same tasks as $\Ch$, where $U'': \chrcompcat \fl \scompcat$ is the forgetful functor.
\end{theorem}
\begin{proof}
As a shorthand notation, we write $
\alpha^n_\Ic := \Ch^{n-1}(\alpha_\Ic) \circ \Ch^{n-2}(\alpha_{\Pi(\Ic)}) \circ\cdots\circ\Ch(\alpha_{\Pi^{n-2}(\Ic)})\circ\alpha_{\Pi^{n-1}(\Ic)},
$ 
and $\alpha^0_\Ic = \alpha_\Ic$. Note that $\Ch^i(\alpha_\Cr)$ is well-defined since $\Ch$ is defined on $\Posf$, see Remark~\ref{R:PosEndo} below. It is routine to check that $\alpha^n_\Ic: \Pi^n(\Ic) \fl \Ch^n(\Ic)$. Consider the following diagram:
\[
\xymatrix@C=10em{
&
\Pi^n(\Ic)
	\ar[d] ^-{\alpha_{\Pi^{n-1}(\Ic)}}
	\ar[dl] _-{\rho_{\Pi^{n-1}(\Ic)}}
\\
\Pi^{n-1}(\Ic)
	\ar[d] _-{\alpha^{n-1}_\Ic}
&
\Ch(\Pi^{n-1}(\Ic))
	\ar[l] ^-{r_{\Pi^{n-1}(\Ic)}}
	\ar[d] ^-{\Ch(\alpha^{n-1}_\Ic)}
\\
\Ch^{n-1}(\Ic)
&
\Ch^n(\Ic).
	\ar[l] ^-{r_{\Ch^{n-1}(\Ic)}}
}
\]
The upper triangle commutes by the hypothesis on $\alpha$, while the lower triangle commutes since $r$ is a natural transformation from $\Ch$ to $Id_{\Posf}$. Furthermore, we have $\Ch(\alpha^{n-1}_\Ic) \circ \alpha_{\Pi^{n-1}(\Ic)} = \alpha^n_\Ic$. By the universal property of the limit, we deduce a unique spectral map $\alpha^\infty_\Ic : \speclim\Ic \fl \Ch^\infty(\Ic)$. Applying Theorem~\ref{thm-carac} suffices to conclude. The argument for $\beta$ is precisely the same, with the roles of $\Pi$ and $\Ch$ reversed. In particular, we use the fact that $\beta$ is a natural transformation between endofunctors of $\scompcat$ (resp.~functors $\chrcompcat\fl\scompcat$) to conclude that the maps $\beta^n_\Cr$ are well-defined.
\end{proof}

\begin{remark}\label{R:Scope}
We believe that Theorem~\ref{Theorem3} can be recovered as a direct consequence of the above theorem. Indeed, the chromatic $IIS$ protocol corresponds to an endofunctor on chromatic simplicial complexes~\cite{GMT15}, and we believe that the appropriate natural transformations exist. However, we leave the full scope of this theorem for future work in which the relationships between chromatic and achromatic protocols and colorless and colored task-solvability are further investigated. {Indeed, in the case of a chromatic protocol, for the semantics to faithfully describe the distributed systems in question, the natural transformation $\alpha$ should be compatible with the ``projecting out'' of process identities, which we do not as of yet understand categorically. }
\end{remark}

\subsection{Spectral space of the achromatic IIS model}\label{SS:SpecSpaceIIS}

It is well-known~\cite[Chap.~4.2]{HKRbook} that the Iterated Immediate  Snapshot protocol, in the context of achromatic distributed computing, corresponds to the barycentric subdivision functor, which is mesh-shrinking. Below, we explicit the construction of the limit space for this protocol for an arbitrary number of initial values, and illustrate them for the special cases of two and three initial values.

Consider the achromatic $IIS$ protocol $(\Ch, b)$, where $\Ch : \scompcat \fl \scompcat$ sends a simplicial complex $C$ to the poset $\Ch(C)$ of its chains and their inclusions, which is also a simplicial complex, and where $b_\Ic : \,\Ic \longrightarrow \dlatt{\Ch(\Ic)}$ sends a simplex $\sigma$ to $\below\{\gamma \in \Ch(C) \mid \Sup\gamma = \sigma\}$.
The maps $b_{\Ch^n(\Ic)}$ send a chain $\gamma \in \Ch^n(\Ic)$ to $\below\{\Gamma \in \Ch^{n+1}(C) \mid \Sup\Gamma = \gamma\}$.
The dual map associated to $\jlift {b}_\Ic$ are maps $m_n : \Ch^{n+1}(\Ic) \rightarrow \Ch^n(\Ic)$ defined by $\Gamma \mapsto \Sup\Gamma$.

\begin{remark}\label{R:PosEndo}
The functor $\Ch$ is actually a functor whose domain of definition is the category $\Posf$ of finite posets. Indeed, it is also known as the \emph{nerve functor}, see for example~\cite{GMT15}.
\end{remark}

The maps $m_n$ thus give us the projective limit system defining the spectral space $\speclimch\Ic$. Using Propositions~\ref{P:ProtSpecSpace} and~\ref{P:MaxPoints} and the specificity of this protocol, we will now describe the limit space $\speclimch \Ic$. 
In particular, we will characterize the downsets of maximal points in terms of the terminal co-dimension of their associated minimal sequences when $\Ic$ is the standard $d$-simplex for some $d\geq 0$. 

First, by Proposition~\ref{P:ProtSpecSpace}, we know that the points of $\speclimch\Ic$ correspond to sequences $(\Gamma)$ such that $\Gamma_n \in \Ch^n(\Ic)$ and $\Gamma_n = \bigvee\Gamma_{n+1}$ for all $n \geq 0$.
Now we divide the points $x\in \real\Ic$ into three classes, depending on the terminal co-dimension of the associated minimal sequences $(\Gamma^x)$. More precisely, given $x \in \real{\Ic}$,
\begin{itemize}
\item $x \in C_1$ if $\codim(\Gamma_n^x)= 0$ for all $n$,
\item $x \in C_3$ if $\exists N$ such that $\forall n\geq N$, $\codim(\Gamma_n^x) = 1$,
\item $x \in C_\infty$ if $\exists N$ such that $\forall n\geq N$, $\codim(\Gamma_n^x) \geq 2$.
\end{itemize}

Before proving Proposition~\ref{P:CherryCount}, we need the following lemma.

\begin{lemma}\label{L:CodimOneCover}
Let $\Ic$ be the standard $d$-simplex. For $\sigma \in \Ch^n(\Ic)$ with $\codim(\sigma) = 1$, then 
\begin{itemize}
\item If $\real\sigma \subseteq \interior{\real\Ic}$, there exist two unique simplices $\tau_1, \tau_2 \in \Ch^n(\Ic)$ such that $\sigma\leq \tau_1, \tau_2$.
\item If $\real\sigma \subseteq \partial\real\Ic$, there exists a unique $\tau\in \Ch^n(\Ic)$, $\sigma\leq\tau$.
\end{itemize}
\end{lemma}

\begin{proof}
We proceed by induction on $n\geq 0$. For $n=0$, we, without loss of generality, write $\sigma = \{v_1, \dots, v_{d}\}$ where $V(\Ic) = \{v_1, \dots, v_{d+1}\}$. The latter is the unique maximal simplex that contains $\sigma$, and clearly $\real\sigma \subseteq \partial{\real\Ic}$. 
For $n \geq 0$, $\sigma\in \Ch^{n+1}(\Ic)$ of co-dimension one and let $\sigma' = m_n(\sigma)$. By the induction hypothesis, there exists unique
	\begin{itemize}
	\item $\tau'_1, \tau'_2 \in \Ch^n(\Ic)$ covering $\sigma'$ and $\real{\sigma'}$ is contained in the interior of $\real\Ic$. Thus $\real{\sigma} \subseteq \real{\sigma'}$ is too.		
Now, since $\tau'_i$ is a $d$-simplex, without loss of generality we write $\{v_{1}, \dots, v_d, v_{d+1,i} \}$ for its set of vertices, and $\sigma' = \{v_1 \dots, v_d\}$. Since $\sigma\in b_n(\sigma')$, we can write $ \sigma = (V_1, \dots, V_{d})$ where $V_1 \leq \dots \leq V_d$ is a maximal chain in $\Pow{\sigma'}$. Up to permutation, we can assume without loss of generality that $V_k = {v_1, \dots, v_k}$. A simplex $\tau \in b_n(\tau'_i)$ is a maximal chain $(U_1, \dots, U_{d+1})$ in $\Pow{\tau'_i}$, its faces being $(U_1, \dots, \hat{U}_i U_{d+1})$, where $\hat{U}_i$ means omitting $U_i$ from the chain. Thus, the only simplex in $b_n(\tau'_i)$ covering $\sigma$ is $\tau_i = \{ V_1, \dots, V_{d}, V_d\cup\{v_{d+1,i}\}\}$.
	\item $\tau' \in \Ch^n(\Ic)$ covering $\sigma'$ and $\real{\sigma}\subseteq \real{\sigma'} \subseteq \partial \real\Ic$. Using the same notations as above for $\sigma, \sigma'$, and writing $\tau' = \{v_{1}, \dots, v_d, v_{d+1} \}$, a similar argument shows that the unique simplex in $b_n(\tau)$ covering $\sigma$ is $\tau = \{ V_1, \dots, V_{d}, V_d\cup\{v_{d+1}\}\}$.
	\end{itemize}
\end{proof}

Now we are ready to characterize the number of elements in each down-set under the specialization order.
\begin{proposition}\label{P:CherryCount}
For $x \in \interior{\real \Ic}$, we have
$
x\in C_i
\iff
\#(\below x) = i
$
where the downset is taken in $\Ch^\infty(I)$ and $i \in \{1,3,\infty\}$. For $x \in \partial\real{\Ic}$, the same holds excepting that $x \in C_3$ if, and only if, $\#(\below x) = 2$.
\end{proposition}

\begin{proof}
We proceed case by case:
\begin{itemize}
\item ($x \in C_1$) In this case $\Gamma^x_n$ is maximal in each $\Ch^n(\Ic)$, so its principal in the specialization order is a singleton.
\item ($x\in C_3$) By Lemma~\ref{L:CodimOneCover}, each $\sigma_n^x$ is covered by either exactly one or two simplices depending on whether $x\in \partial{\real\Ic}$ or not, respectively. Since the $m_n$ are order and dimension preserving, these form sequences $(\tau)$ or $(\tau_1)$ and $(\tau_2)$, respectively. Since each $\tau_n$ or the $(\tau_i)_n$ are unique cover of $\sigma_n^x$, these are the unique sequence above $(\sigma^x)$ in the point-wise order, which is the order-dual of the specialization.
\item ($x\in C_\infty$) Let $N$ so that $\codim(\sigma_n^x) \geq 2$ for all $n \geq N$. There exists $\tau_N$ of codimension one such that $\sigma_N^x \leq \tau_N$. The latter is covered by two maximal simplices $\tau_N^1$ and $\tau_N^2$. By a similar argument to that given in Lemma~\ref{L:CodimOneCover}, we show that $\sigma_{N+1}^x$ is below by at least four maximal simplices $\tau_{N+1}^{11}, \tau_{N+1}^{12}, \tau_{N+1}^{21}, \tau_{N+1}^{22}$  in $\Ch^{N+1}(\Ic)$, where $\tau_{N+1}^{i1}, \tau_{N+1}^{i2} \in b_N(\tau_N^i)$. By induction we thus conclude that $\sigma_{N+k}^x$ is covered by at least $2^{k+1}$ maximal simplices in $\Ch^{N+k}(\Ic)$, and that we have constructed $2^{k+1}$ maximal sequences that are above $(\sigma_N^x, \dots \sigma_{N+k}^x)$ in the point-wise order on $\prod_{i=N}^{N+k}\Ch^i(\Ic)$. This shows that taking $k$ to infinity, we obtain an infinite amount of sequences above $(\sigma_n^x)$.
\end{itemize}
\end{proof}

Below, we illustrate the spectral spaces $\speclimch\Ic$ when $\Ic$ is the $d$-simplex for $d=1,2$. For the $1$-simplex, $\speclimch\Ic$ is the space pictured below in blue. We have included the second barycentric subdivision of the standard $1$-simplex above it for reference. Firstly, we observe that the terminal co-dimension of every sequence is either $0$ or $1$, so each point will either have $0$ or $2$ points below it, respectively. The subspace of maximal points is isomorphic to the real interval $[0,1]$, this being the geometric realization of $\Ic$. At each point $x$ of the form $\frac{k}{2^n}$, for $n\geq 1$ and $0<k<2^n$, \ie those that correspond to a vertex in one of the subdivisions, we have two points below $x$ in the specialization order, as is the case for $x_3$. Extremal points like $x_2$ have terminal co-dimension $1$, but only have one element below them since they are in the border of the realization. All other points in the interval have no elements below them. This is case for $x_1=\frac{1}{3}$, for example.

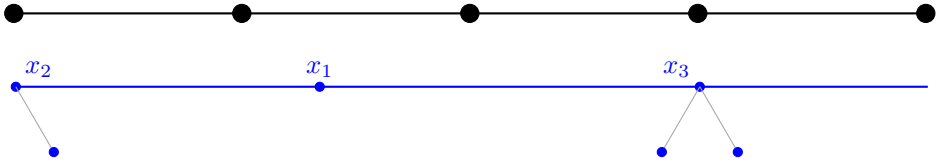
\begin{center}
\begin{tikzpicture}[scale=4]

\coordinate (A) at (0, 0);
\coordinate (B) at (3, 0);

\coordinate (ABm) at ($(A)!0.5!(B)$);
\coordinate (Am) at ($(A)!0.5!(ABm)$);
\coordinate (Bm) at ($(ABm)!0.5!(B)$);
\coordinate (x1) at ($(Am)!0.333!(ABm)$);

\draw[thick] (A) -- (B);

\foreach \point in {A, B, ABm, Am, Bm} {
    \filldraw[black] (\point) circle (0.03);
}
\end{tikzpicture}
\vskip10pt
\begin{tikzpicture}[scale=4]

\coordinate (A) at (0, 0);
\coordinate (B) at (3, 0);

\coordinate (ABm) at ($(A)!0.5!(B)$);
\coordinate (Am) at ($(A)!0.5!(ABm)$);
\coordinate (Bm) at ($(ABm)!0.5!(B)$);
\coordinate (x1) at ($(Am)!0.333!(ABm)$);

\draw[thick,blue] (A) -- (B);

\node[above right] at (A) {{\color{blue}$x_2$}};
\node[above left] at (Bm) {{\color{blue}$x_3$}};
\node[above] at (x1) {{\color{blue}$x_1$}};

\foreach \point in {A, Bm, x1} {
    \filldraw[blue] (\point) circle (0.015);
}
\def\raylen{0.25}

\foreach \P in {Bm} {
    \foreach \angle in {-60, -120} {
        \path (\P) ++({\angle}:\raylen) coordinate (End);
        \draw[gray!70] (\P) -- (End);
        \filldraw[blue] (End) circle (0.015);
    }
}

\foreach \P in {A} {
    \foreach \angle in {-60} {
        \path (\P) ++({\angle}:\raylen) coordinate (End);
        \draw[gray!70] (\P) -- (End);
        \filldraw[blue] (End) circle (0.015);
    }
}

\end{tikzpicture}
\end{center}

For $d=2$, we have pictured the first barycentric subdivision on the left, and the limit space $\speclimch\Ic$ on the right. The point $x_1$ is an example of a point such that $\sigma^{x_1}_n$ is of dimension $2$ for all $n$, thus there are no elements below it. Points whose associated minimal sequences are of co-dimension $1$ have exactly two elements below them if they are interior, like $x_2$, or one element below them if not, as is the case for $x_3$. Finally, points like $x_4$ and $x_5$ have a poset of points below them of infinite width and depth of two. 

\[
\raisebox{16pt}{
\begin{tikzpicture}[scale=5]

\coordinate (A) at (0, 0);
\coordinate (B) at (1, 0);
\coordinate (C) at (0.5, 0.866); %

\coordinate (ABm) at ($(A)!0.5!(B)$);
\coordinate (BCm) at ($(B)!0.5!(C)$);
\coordinate (CAm) at ($(C)!0.5!(A)$);

\coordinate (G) at ($ (A)!.3333!(B) + 0.3333*(C) - 0.3333*(A) $);

\coordinate (x2) at ($ (ABm)!0.33!(G)$);
\coordinate (x1) at ($ (A)!.4!(B) + 0.5*(C) - 0.3333*(A) $);
\coordinate (x4) at ($ (BCm)!0.66!(B)$);

\fill[blue!10] (A) -- (B) -- (C) -- cycle;

\draw[thick] (A) -- (B) -- (C) -- cycle;

\draw (A) -- (ABm);
\draw (B) -- (BCm);
\draw (C) -- (CAm);

\draw (ABm) -- (G);
\draw (BCm) -- (G);
\draw (CAm) -- (G);

\draw (A) -- (G);
\draw (B) -- (G);
\draw (C) -- (G);

\foreach \point in {A, B, C, ABm, BCm, CAm, G} {
    \filldraw[black] (\point) circle (0.02);
}

\end{tikzpicture}
}
\qquad\qquad
\begin{tikzpicture}[scale=5]

\coordinate (A) at (0, 0);
\coordinate (B) at (1, 0);
\coordinate (C) at (0.5, 0.866); %

\coordinate (ABm) at ($(A)!0.5!(B)$);
\coordinate (BCm) at ($(B)!0.5!(C)$);
\coordinate (CAm) at ($(C)!0.5!(A)$);

\coordinate (G) at ($ (A)!.3333!(B) + 0.3333*(C) - 0.3333*(A) $);

\coordinate (x2) at ($ (ABm)!0.33!(G)$);
\coordinate (x1) at ($ (A)!.4!(B) + 0.5*(C) - 0.3333*(A) $);
\coordinate (x4) at ($ (BCm)!0.66!(B)$);

\fill[blue!10] (A) -- (B) -- (C) -- cycle;

\draw[thick,blue] (A) -- (B) -- (C) -- cycle;

\node[above left] at (x1) {{\color{blue}$x_1$}};
\node[below right] at (x2) {{\color{blue}$x_2$}};
\node[right] at (x4) {{\color{blue}$x_3$}};
\node[above left] at (A) {{\color{blue}$x_5$}};
\node[left] at (G) {{\color{blue}$x_4$}};

\foreach \point in {x1, G, x2, x4, A} {
    \filldraw[blue] (\point) circle (0.01);
}

\def\raylen{0.08}
\def\smallray{0.04}
\def\angleoffset{20} 

\foreach \P in {x4} {
    \foreach \angle in {-120} {
        \path (\P) ++({\angle}:\raylen) coordinate (End);
        \draw[gray!70] (\P) -- (End);
        \filldraw[blue] (End) circle (0.01);
    }
}

\foreach \P in {x2} {
    \foreach \angle in {-120,-60} {
        \path (\P) ++({\angle}:\raylen) coordinate (End);
        \draw[gray!70] (\P) -- (End);
        \filldraw[blue] (End) circle (0.01);
    }
}

\foreach \P in {A, G} {
    \foreach \angle in {-150,-120,-60,-30} {

        \path (\P) ++({\angle}:\raylen) coordinate (End);
        \draw[gray!70] (\P) -- (End);
        \filldraw[blue] (End) circle (0.01);

        \foreach \delta in {-\angleoffset, \angleoffset} {
            \path (End) ++({\angle + \delta}:\smallray) coordinate (SubEnd);
            \draw[gray!50] (End) -- (SubEnd);
            \filldraw[blue] (SubEnd) circle (0.01);
        }
    }
}
\path (G) ++({-90}:.1) coordinate (dotsx4);
\node at (dotsx4) {{$\scriptstyle\dots$}};

\path (A) ++({-90}:.1) coordinate (dotsx5);
\node at (dotsx5) {{$\scriptstyle\dots$}};

\end{tikzpicture}
\]

\subsection{Protocols as functors with natural transformations}\label{A:ProtocolsAsFunctors}

Here we justify our functorial approach to achromatic protocols, describing how the Yoneda extension allows us to glue together local information about the action of the distributed protocol and the associated carrier maps, thereby extending to an endofunctor on the category of simplicial complexes with an associated natural transformation as defined in Section~\ref{SS:ProtEndo}. The argument for chromatic protocols is similar.

As explained in Section~\ref{SS:ProtEndo} and in the sections above in the appendix, we consider {content-neutral, round-based full-information} protocols. 
What this means concretely is that given a global state, \ie a set $\sigma = \{v_1, \dots, v_n\}$ of local states, a protocol $\Pi$ returns the collection of possible global states $\sigma_1, \dots, \sigma_{k_n}$ 
after one round of execution. We denote by $\Pi(\sigma)$ the simplicial complex spanned by these sets. Considering all $\sigma$, this forms the protocol complex. 
Since $\Pi(\sigma)$ contains the possible states obtainable after one round of computation, we have $\bigcup_{\sigma'\subseteq\sigma} \Pi(\sigma') = \Pi(\sigma)$. For $\sigma_1, \sigma_2 \subseteq \sigma$, operationally, $\Pi(\sigma_1) \cap \Pi(\sigma_2)$ consists of all reachable states in which no process can know whether inputs were from $\sigma_1$ or $\sigma_2$.
The only processes which can participate in such an execution are those having values in $\sigma_1\cap\sigma_2$, because the others are able to distinguish between these two initial configurations.
These executions are precisely those whose global states after one round of execution are in $\Pi(\sigma_1 \cap \sigma_2)$, so we have
\[
\Pi(\sigma_1) \cap \Pi(\sigma_2) = \Pi(\sigma_1 \cap \sigma_2).
\]
Thus, given a subset $\sigma' \subseteq \sigma$, we have $\Pi(\sigma') \subseteq \Pi(\sigma)$. Sub-simplices of $\sigma$ are therefore mapped to sub-complexes of $\Pi(\sigma)$, defining an order-preserving map $\pi_\sigma : \Pow\sigma \fl \dlatt{\Pi(\sigma)}$. 
  Because we assume that our protocol is full-information, it is not possible that all executions contain as much information as some execution from a strict subset of processes. Formally, this means that
it is not possible that for all $\tau \in \Pi(\sigma)$, $\bigcap_{\sigma'\subseteq \sigma \\ \tau \in \Pi(\sigma')} \sigma' \subsetneq \sigma$.
Equivalently, there must exist some execution in which no information is lost, and thus there exists some global state $\tau \in \Pi(\sigma)$ such that $\sigma = \bigcap_{\sigma'\subseteq \sigma \\ \tau \in \Pi(\sigma')} \sigma'$. In particular, this means that the join-lifts of the $\pi_\sigma$ are injective.
Finally, since $\Pi$ is content-neutral, the actual values $v_1, \dots, v_n$ do not affect the outcome $\Pi(\sigma)$, meaning that we may consider two inputs $\sigma, \sigma'$ equivalent when they are of the same cardinality. For this reason, the category of finite sets and set maps provide an appropriate description of the system states.

Let $f:\sigma \fl \sigma'$ be a mapping of sets. Since $\Pi$ is full information, each of the vertices of $\Pi(\sigma)$ and $\Pi(\sigma')$ consist of views containing the values of $\sigma$ and $\sigma'$, respectively. Replacing each value $v \in \sigma$ by $f(v)$ in the vertices of $\Pi(\sigma)$ thus yields a vertex of $\Pi(\sigma')$. This yields a simplicial map since $\Pi(\sigma'\cap f[\sigma]) = \Pi(\sigma')\cap\Pi(f[\sigma])$ is a subcomplex. Notice that if $f$ is the canonical inclusion $\sigma\subseteq\sigma'$, this mapping yields the canonical inclusion $\Pi(\sigma) \subseteq \Pi(\sigma')$.

Summing this up, and denoting by $\sigma_n = \{1, \dots, n\}$ the standard $n$-simplex, by $\Delta_n = \Pow\sigma\setminus\{\emptyset\}$ the standard $n$-simplex viewed as a complex, and by $\simpcat$ the category of standard $n$-simplices and set-maps between them, we have a functor 
$\Pi : \simpcat \fl \scompcat$, 
along with a natural transformation $\pi : U\circ\yo \Rightarrow V\circ\mathcal{D}\circ\Pi$, where $\yo : \simpcat \fl \scompcat$ is the Yoneda embedding $\sigma_n \mapsto \Delta_n$, and $U : \scompcat \fl \Posf$ and $V : \DL \fl \Posf$ are forgetful functors. Moreover, each component of $\pi_{\sigma_n}$ is a strict carrier map which is moreover effectively surjective, \ie the join of its images is the full complex $\Pi(\sigma_n)$. 

Now we describe how to extend this information to the definition provided in Section~\ref{SS:ProtEndo} using the Yoneda extension. Given a simplicial complex $\Cr$ and denoting by $D_\Cr$ the diagram in $\simpcat$ of inclusions of simplices of $\Cr$, we have $\Cr = \colim{(\yo(D_\Cr))}$. Denoting by $\Pi(D_\Cr)$ the image of this diagram under $\Pi$, and by $\Pi(\Cr)$ its colimit, we have extended $\Pi$ to an endofunctor of $\scompcat$. Indeed, we have thereby defined an assignment on objects, and, given a simplicial map $f:\Cr \fl \Cr'$, it comes from a map $V_\Cr \fl V_{\Cr'}$ which defines a natural transformation from $D_\Cr$ to $D_{\Cr'}$. Indeed, for any simplex $\sigma = \{v_1, \dots, v_k\}$, we have a map of sets $f_{\mid\sigma} : \sigma \fl f[\sigma]$, where $f[\sigma] = \{f(v_1), \dots, f(v_k)\}$.
 Applying $\Pi$, we see that all of this entails that $\Pi(\Cr')$ is a co-cone for $\Pi(D_\Cr)$, which, by the universal property of the colimit, provides a unique map $\Pi(f) : \Pi(\Cr) \fl \Pi(\Cr')$.
 
 The argument extending the natural transformation is similar. Indeed, for a complex $\Cr$, noticing that forgetful functors preserve colimits, we have that $\Cr$ is the colimit in $\Posf$ of the diagram $U(\yo(D_\Cr))$, and since $\mathcal D$ also preserves colimits, we have that $\dlatt{\Pi(\Cr)}$ is the colimit in $\Posf$ of the diagram $V(\dlatt{\Pi(D_\Cr)})$. By construction, the appropriate components of $\pi$ are a natural transformation between $U(\yo(D_\Cr))$ and $V(\dlatt{\Pi(D_\Cr)})$, thereby providing a unique $\pi_\Cr : U(\Cr) \fl V(\dlatt{\Pi(\Cr)})$ by the property of the colimit. Combining this with the arguments used to prove that $\Pi$ is functorial on $\scompcat$, we deduce the commutativity of naturality squares. 
 Finally, preservation of intersections, \ie strictness, and the join of images being the full complex, \ie essential surjectivity, are preserved under these constructions. Moreover, since each of the join-lifts of the maps in the colimit are injective, and all the maps involved in each of the diagrams $U(\yo(D_\Cr))$ (and thereby in $U(\dlatt{\yo(D_\Cr)})$) and $V(\dlatt{\Pi(D_\Cr)})$ are inclusions, we conclude that the map $\jlift\pi_\Cr$ is also injective.
 
 Thus, we have shown that from a full-information content-neutral protocol, we deduce a pair $(\Pi, \pi)$, where $\Pi : \scompcat \fl \scompcat$ is a functor and $\pi$ is a natural transformation between $U: \scompcat \fl \Posf$, the forgetful functor, and $\Dr\circ\Pi : \scompcat \fl \Posf$ whose components are strict and essentially surjective carrier maps, and whose join-lifts are injective.

 \begin{remark}
   There are no mention of functors or functorial presentation in the
   reference book \cite{HKRbook}, however, while generalizing their
   results, they introduce conditions (like border-consistency for
   general subdivision protocol in Def.3.6.5) that seems to imply
   functoriality of the models under consideration.
 \end{remark}

\section{Distributed Models of Computation}

In this section we detail the precise relation of our functorial
presentation with standard distributed models of computation.  
The first subsections are intended for people with few or no knowledge about combinatorial topology for distributed computing.
The last section explains how  Yoneda extension enables to get a fully functorial presentation for some classical distributed models.

\subsection{About combinatorial topology encoding for distributed computing}
\label{sec:gencombtopo}
We present shortly the way many distributed computations can be encoded using
simplicial complexes. Since this is not so simple to summarize
entirely, we refer the interested reader to the best reference which
is the book \cite{HKRbook}. There is also a short introduction in \cite{DCcolumn}.
As presented previously in Section~\ref{S:OFIFunc}, it is possible to describe a distributed
system using simplicial complexes.  Before presenting in more detail how
to precisely encode for various models, we explain why  round-based simplicial
descriptions are actually sufficient to investigate executions for
a large family of distributed systems.

A distributed system is constituted of
processes and shared objects. Processes access and/or modify shared
objects by using primitives. A primitive is called by a process as a
local function, that manipulates the given shared object. The way
simultaneous calls to a given object behave depends upon the
properties (or specification) of the object.
An object can be one-shot (used only once by some processes) or long lived, that is accessed and updated repeatedly (\eg a shared stack).
A global state in a distributed system is therefore described by the tuple of local states of
participating processes together with the internal state of shared
objects used by these processes. What has been discovered is that many
shared objects (like shared registers) reused by the processes could
be replaced with one-shot objects (like snapshots) with equivalent
wait-free computability. Moreover, it was also proved that these
one-shot objects could be equivalently used in a layered way, that is,
as if the system was synchronous. In the end, this means that we need
to consider only round-based protocols and when using full information
algorithms, there is not need to include the (one-shot) state of
global objects to describe in integrality the global state of the
system. Therefore a sequence of the tuples of local states is
sufficient to describe the possible evolution of many distributed
systems. This culminates in the so called message adversary models,
where the message are to be delivered (or definitely lost) is the same
round (see below).

In short, by using algorithmic reductions, it was shown that what is
computable by asynchronous wait-free\footnote{the number of crash failures is unlimited} shared memory systems is the same
as what is computable by some specific synchronous fault-free
message-passing systems, like the one in Def.~\ref{def-iis} below.

\subsection{Messages Adversaries}

\label{sec:mashort}
Now we present precise encodings for different distributed models,
presented as message adversaries (this terminology was coined in \cite{messadv} but the model appeared with various names beforehand).

As specific content-neutral, round-based, full-information models, we consider the general
message adversary models.
For $d\in\NN$, we denote by $[d]=\{0,..,d\}$ the set of
processes. We use standard directed graph (or digraph) notations : given $G$, $V(G)$ is
the set of vertices, $A(G)\subseteq V(G)\times V(G)$ is the set of arcs.
A \emph{dynamic graph} \dG is a sequence $G_1,G_2,\cdots,G_r,\cdots$ where
  $G_r$ is a directed graph with vertices in $[d]$. We also denote by $\dG(r)$ the digraph $G_r$.
  A \emph{message adversary} is a set of dynamic graphs.
Intuitively, the arcs $(u,v)$ of $\dG(r)$ describe the transmission of messages from $u$ to $v$ at
round $r$. If there is no arc from $u$ to $v$, no information is transmitted. Note that $u$ is not a priori aware whether a given broadcast message has been received. A formal definition of an execution under a dynamic graph is
given in Section~\ref{execution}.

We will use the following standard notations in order to more easily describe
our message adversaries \cite{PPinfinite}. A dynamic graph is
seen as an infinite word over the alphabet $\mathcal G_d$, set of all graphs with vertices $[d]$.%
  Given $U\subseteq\mathcal G_d$, $U^*$ is the set of all finite
  sequences of elements of $U$, $U^\omega$ is the set
  of all infinite sequences, and $U^\infty = U^* \cup U^\omega.$
An message adversary \ma is said to be
\emph{compact} (in the sense of the profinite topology on infinite
words~\cite{PPinfinite}) if for any word $\dG\in\mathcal G_d^\omega$, $\dG_{\mid r}\in Pref_r(\ma)$ for all $r\in\NN$ implies
$\dG\in\ma$, where $Pref_r(\ma)$ is the set of prefixes of length $r$ of \ma.

\subsection{Execution of a distributed algorithm}
\label{execution}
Given a message adversary $\ma$ and a set of initial configurations \In,
we define what is an execution of a given algorithm \algo subject to $\ma$ with initialization \In.
An execution is constituted of an initialization step, and a (possibly infinite) sequence of rounds of
messages exchanges and corresponding local state updates.
When the initialization is clear from the context, we will use
\emph{scenario} and \emph{execution} interchangeably.

An execution of an algorithm \algo under scenario $w\in \ma$ and initialization $\iota\in\In$
is the following. This execution is denoted $\iota.w$.
First, $\iota$ affects the initial state to all
processes of $[n]$. Then the system progresses in rounds.

A round is decomposed in 3 steps : sending, receiving, updating the
local state.  At round $r\in\mathbb N$, messages are sent by the processes
using the \texttt{SendAll()} primitive. The fact that the corresponding
receive actions, using the \texttt{Receive()} primitive, will be successful depends on $G=w(r)$,
$G$ is called the \emph{instant graph} at round $r$.

Let $p,q\in\Pi$. The message sent by $p$ is received by $q$ on the
condition that the arc $(p,q)\in A(G)$.  Then, all processes update
their state according to the received values and \algo.  Note that it
is assumed that $p$ always receives its own value, that is $(p,p)\in
A(G)$ for all $p$ and $G$. However, in examples, this might be
implicit for clarity and brevity.

\medskip
Let $w\in\ma, \iota\in\In$. Given $u\in Pref(w)$, we denote by $\mathbf s_p(\iota.u)$ the
state of process $p$ at the $len(u)$-th round of the algorithm \algo
under scenario $w$ with initialization $\iota$. This means  that
$\mathbf s_p(\iota.\varepsilon) = \iota(p)$ represents the initial state of $p$ in $\iota$, where
$\varepsilon$ denotes the empty word.

\subsection{Decision of a distributed algorithm}
\label{S:decision}
When presented in a full-information way, a distributed algorithm
actually reduces to a decision function that should state when output
a decision and which decision value according to the local state.

This can be seen a function from the vertices of the protocol complex
to the vertices of the output complex. Of course, when the distributed
model is anonymous, one has to check that the function does not
actually use the knowledge of which process it is. In this paper,
since we consider only colorless tasks, this is actually irrelevant.

\subsection{Classical examples of message adversaries}
We show how standard distributed models are described in
this framework.
We fix an integer $d$.

\begin{example}
  Consider a message passing system with $d+1$ processes where, at each round, all messages could be
  lost. The associated message adversary is $\mathcal G_{d}^\omega$. For two processes \b and \n, we have
  $\mathcal G_{1}=\{\lblanc,\lok,\lnoir,\lall\}$.
\end{example}

Given a graph $G$, we denote by
$In_G(a) = \{ b\in V(G) \mid (b,a)\in A(G)\}$ the set of incoming vertices of $a$ in $G$.
A graph $G$ has the \emph{containment property} if for all $a,b\in
V(G)$,
\begin{itemize}
\item $a\in In_b$ or $b\in In_a$;
\item $In_G(a)\subseteq In_G(b)$ or $In_G(b)\subseteq In_G(a)$.
\end{itemize}

\begin{definition}[Chap.~14.2 in \cite{HKRbook}]\label{def-is}
We set
$S_d = \{G\in\mathcal G_d\mid G$ has the
Containment property$\}$.
The Iterated Snapshot message adversary for $d+1$ processes is the
message adversary $IS_d=S_d^\omega$.
\end{definition}

We say that a graph $G$ has the \emph{Immediacy Property} if for all $a,b,c\in V(G)$, $(a,b),
(b,c)\in A(G)$ implies that $(a,c)\in A(G)$.

\begin{definition}[Chap.~4.3 and 14 in \cite{HKRbook}]\label{def-iis}
  We set
  $ImS_d = \{G\in\mathcal G_d\mid G$ has the Immediacy and
 \mbox{Containment} properties $\}$.
The Iterated Immediate Snapshot message adversary for $d+1$ processes is the
message adversary $IIS_d=ImS_d^\omega$.
\end{definition}

There exist variants of these models, the chromatic and achromatic variants. When, with the \texttt{Receive} primitive, a process is able to distinguish messages with the same content arriving from different processes, this is the (standard) chromatic version. This amounts to processes having unique identities.
When the value returned by the \texttt{Receive} primitive is actually a set of the broadcast values, which means that  a process is not able to distinguish messages with the same content arriving from different processes -- including itself  --, this is the achromatic variant. Note that any homonymous model (processes have labels that may not be unique) can also be expressed in that setting.

The one round complex of the chromatic (resp. achromatic) $IIS$ protocol is the standard chromatic subdivision, \cite[Chap.~8]{HKRbook} and below, (resp. barycentric subdivision, \cite[Chap.~5]{HKRbook}). The protocol complex for the colored $IS$ model is described in Appendix~\ref{sec-protc}.
Finally, the corresponding functors can be recovered by a Yoneda extension argument, see Appendix~\ref{A:ProtocolsAsFunctors} below, or~\cite[Section~4.1]{catdist-x} for a similar argument.
It is currently not known precisely under which conditions a given message adversary defines a functorial presentation.

\subsection{Protocol complexes of some message adversaries}

\label{sec-protc}

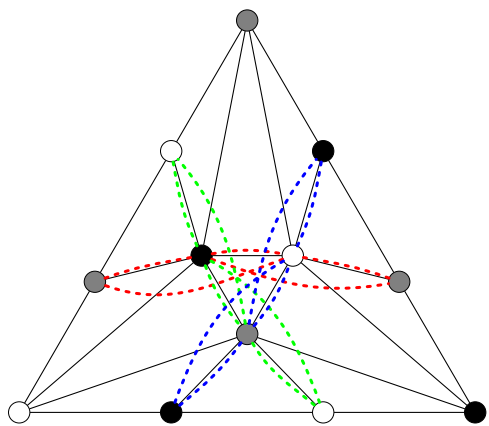
\begin{figure}[h]
\centering
\begin{tikzpicture}[scale=.6]

      \node[proc,g,label=north:] (0) at (5.0, 8.660254037844386) {};
      \node[proc,b,label=south:] (1) at (0, 0) {};
      \node[proc,n] (2) at (10, 0) {};

      \node[proc,g,label=left:] (3) at (5.0, 1.7320508075688772) {};
      \node[proc,b] (4) at (6.0, 3.4641016151377544) {};
      \node[proc,n] (5) at (4.0, 3.4641016151377544) {};
      
      \draw (3) -- (4); 
      \draw (4) -- (5); 
      \draw (5) -- (3);

      \node[proc,n,label=south:] (6) at (3.3333333,0) {}; 
      \node[proc,b] (7) at (6.6666666,0) {};
      
      \draw (1) -- (6); 
      \draw (6) -- (7); 
      \draw (7) -- (2);

     \node[proc,b] (8) at(3.333,5.773) {};
     \node[proc,g] (9) at(1.66,2.886) {};
     
     \draw (0) -- (8); 
     \draw (8) -- (9); 
     \draw (9) -- (1);

     \node[proc,n] (10) at(6.666,5.773) {};
     \node[proc,g] (11) at(8.3333,2.886) {};
     
     \draw (0) -- (10); 
     \draw (10) -- (11); 
     \draw (11) -- (2); 

     \draw (0) -- (4); 
     \draw (0) -- (5);
     \draw (1) -- (5); 
     \draw (1) -- (3); 
     \draw (2) -- (3); 
     \draw (2) -- (4); 

     \draw (3) -- (6); 
     \draw (3) -- (7); 
     \draw (4) -- (10); 
     \draw (4) -- (11); 
     \draw (5) -- (8); 
     \draw (5) -- (9);

     \tikzstyle{dotted}= [dash pattern=on \pgflinewidth off 1mm, line cap= round]  
     
     \draw[red,dotted,line width = 0.4mm] (9) to[out=20,in=190] (5);
     \draw[red,dotted,line width = 0.4mm] (4) to[out=200, in=-20] (9);
     \draw[red,dotted,line width = 0.4mm] (5) to[out=10,in=170] (4);

     \draw[red,dotted,line width = 0.4mm] (5) to[out=-20,in=190] (11);
     \draw[red,dotted,line width = 0.4mm] (4) to[out=-10,in=160] (11);
     
     \draw[green,dotted,line width = 0.4mm] (5) to[out=-30,in=110] (7);
     \draw[green,dotted,line width = 0.4mm] (3) to[out=-60,in=140] (7);
     \draw[green,dotted,line width = 0.4mm] (5) to[out=-70,in=130] (3);

     \draw[green,dotted,line width = 0.4mm] (5) to[out=115,in=-80] (8);
     \draw[green,dotted,line width = 0.4mm] (3) to[out=100,in=-50] (8);
     
     \draw[blue,dotted,line width = 0.4mm] (4) to[out=65,in=-100] (10);
     \draw[blue,dotted,line width = 0.4mm] (3) to[out=80,in=-130] (10);
     \draw[blue,dotted,line width = 0.4mm] (3) to[out=50,in=-110] (4);
     
     \draw[blue,dotted,line width = 0.4mm] (4) to[out=-150,in=70] (6);
     \draw[blue,dotted,line width = 0.4mm] (3) to[out=-125,in=40] (6);
      
  \end{tikzpicture}
  
  \caption{A representation of the complexes corresponding to a round of 
  the chromatic Iterated Snapshot and Iterated Immediate Snapshot model for 3 processes. \label{fig:IS_2-colored}}
\end{figure}

In Figure \ref{fig:IS_2-colored}, we present the protocol complex for one round of $IIS_2$ and one round of $IS_2$ in the chromatic case. The triangles with black edges belong to $\Pi_{IIS}$, and we represented in colored dashed lines 
the additional simplices corresponding to instant graphs of $S_2$
that are not in $ImS_2$. 
We emphasize that, in the figure, when the dashed line is very close to a black line, it actually represents the same simplex of dimension 1.  This protocol complex can
be iterated, however the corresponding complex is not as
easily drawn as the one for the $IIS$ model.

\end{document}